\documentclass[12pt]{article}
\usepackage{epsfig}
\normalbaselines
\newcommand{\be}{\begin{eqnarray}}
\newcommand{\ee}{\end{eqnarray}}

\def\ll#1{\left#1}
\def\r#1{\right#1}
\def\fr{\frac{1}{2}}

\def\mref#1{(\ref{#1})}

\def\p{\partial}

\def\bd{\begin{displaymath}}
\def\ed{\end{displaymath}}

\def\ba#1{\begin{array}{#1}}
\def\ea{\end{array}}
\def\nn{\nonumber}

\begin{document}

\pagestyle{empty}

\begin{center}

{\LARGE\bf The semiclassical small-$\hbar$ limit of loci of roots of fundamental solutions for polynomial potentials\\[.5cm]}

\vskip 60pt

{\large {\bf Stefan Giller}}

\vskip 18pt

Jan Dlugosz Academy in Czestochowa\\
Institute of Physics\\ul. Armii Krajowej 13/15, 42-200 Czestochowa, Poland\\
e-mail: stefan.giller@ajd.czest.pl
\end{center}

\vspace{3 cm}

\begin{abstract}In this paper a description of the small-$\hbar$ limit of loci of zeros of fundamental solutions for
polynomial potentials is given. The considered cases of the potentials are bounded to the ones which provided us with
simple turning points only. Among the latter potentials still several cases of Stokes graphs the potentials
provide us with are distinguished, i.e. the general non-critical Stokes graphs, the general critical ones but with only
single internal Stokes line and the Stokes graphs corresponding to arbitrary multiple-well real even degree polynomial
potentials with internal Stokes lines
distributed on the real axis only. All these cases are considered in their both versions of the quantized and not
quantized $\hbar$. In particular due to the fact that the small-$\hbar$ limit is semiclassical it is shown that loci
of roots of fundamental solutions in the cases considered are collected along Stokes lines. There are
infinitely many roots of fundamental solutions on such lines escaping to infinity and a finite number of them on internal
Stokes lines.

\vskip 36pt

PACS number(s): 03.65.-W , 03.65.Sq , 02.30.Lt , 02.30.Mv


\end{abstract}

\newpage

\pagestyle{plain}

\setcounter{page}{1}

\section{Introduction}

\hskip+1.5em In this paper we are continuing in the case of $\hbar\to 0$ a description of loci of zeros of fundamental
solutions to Schr{\"o}dinger equation (SE) given in our
previous paper \cite{11} for the case of the high energy limit $E\to\infty$. The case $\hbar\to 0$ has been considered recently by
Hezari \cite{9} who investigated in this limit the
problem of complex zeros of eigenfunctions of SE with real polynomial potentials of even degree, while the energy
parameter $E$ was kept fixed and $\hbar$ was quantized.
In fact the two cases, i.e. the
energy quantized while the Planck constant kept fixed and the energy kept fixed but the Planck constant is quantized
have two
different semiclassical limits for the quantized parameters, i.e. the high energy limit and the small $\hbar$-limit lead
each to different behaviour of the corresponding Stokes graphs and to different sets of eigenfunctions. Nevertheless since
both these limits are of the same semiclassical nature at least mathematically it is not surprising that
Hezari's results on complex zeros eigenfunctions problem are similar to those of Eremenko {\it et al} \cite{12}. Note
also the respective studies done by Martinez-Filkenstein {\it et al} \cite{15} and Zelditch \cite{14} as well as by
Delabaere {\it et al} \cite{13}.

In this paper we would like to generalize the results of Hezari. Particularly we would like:
\begin{enumerate}
\item to establish for a general polynomial
potential theorems on zeros distributions of fundamental solutions in the limit $\hbar\to 0$ which are
analogues of the corresponding theorems established in our previous paper \cite{11} for $E\to\infty$
\item to get corresponding results of this limit for a general real double-well (D-W) polynomial
potential of even degree and its multiple-well generalization.
\end{enumerate}

In the first case a general result has been obtained
for polynomials providing Stokes graphs with simple turning points only and at most with one inner Stokes line while
in the second case we have considered
initially a real D-W polynomial potential of tenth degree but of a form that allows us for a simple generalization of the obtained
results to general multiple-well real polynomial potentials of the even degree.

The paper is organized as follows.

In sec.2 the main problem of this paper is formulated and two general theorems are obtained for the limit loci of zeros of
fundamental solutions: for a general non-critical Stokes graph corresponding to a polynomial potential and for such a
graph with a unique internal Stokes line.

In sec.3 the fundamental solutions which sectors are linked by a unique internal Stokes line are quantized and changes
caused by the quantization in the limit loci of their zeros are notified accordingly.

In sec.4 a double-well potential is considered and the limit distributions of zeros of two fundamental solutions
vanishing in the infinities of the real axis are found.

In sec.5 the same solutions are quantized (matched to each other) and changes caused by the quantization in the limit loci
of their zeros are investigated.

In sec.6 the symmetric case of the double-well potential is considered together with the limit loci of zeros of the same
pair of quantized fundamental solutions.

In sec.7 simple generalizations of the results from the sections 5-7 to multiple-well potentials are done.

In sec.8 we summarize the results of the paper.

\section{The non-quantized cases of the limit $\hbar\to 0$}

\hskip+2em It is worth to note that the limit $\hbar\to 0$ can be considered in fact for the not quantized Planck
constant and this case is even much easier
to handle at least as long as all turning points are simple. In the last cases our method developed in the previous paper
\cite{11} can
be applied equally well despite the fact that
positions of the roots can be now arbitrary. The method can certainly be used safely for the non critical SG's as well as
for the critical ones with a single internal Stokes line.

To begin with consider the Schr{\"o}dinger equation for the case:
\be
\phi''(z)-\lambda^2(P_n(z)-E)\phi(z)=0
\label{1}
\ee
with $P_n(z)=a_nz^n+...+a_1z,\;a_n\neq 0,\;n\geq 1$, and $\lambda^2=\frac{2m}{\hbar^2},\;\lambda>0$, and all $a_n,\;n\geq 1$,
are complex and $\lambda$-independent.

By assumption all roots of the equation:
\be
P_n(z)-E=0
\label{1b}
\ee
are simple.

To make our paper self-contained we shall equip it with all necessary notions allowing us to introduce the fundamental
solutions (FS's) to eq.\mref{1} and to discuss their main properties utilized in the paper despite the fact that all
these necessary notion can be found also in our earlier papers \cite{11,3}.

While considering cases of double- and multiple-well potentials we shall limit ourselves to the real and positive
$\lambda$ it is worthwhile to consider it in this section as a complex parameter.

First with eq.\mref{1} and the potential $P_n(z)$ the set of all Stokes lines (SL's) called Stokes graph (SG) is associated.
The SL's are defined by the conditions:
\be
\Re\ll(\lambda \int_{z_i}^{z} \sqrt{W_n(\xi)}d\xi\r) = 0,\;\;\;\;\;\;\;\;i=1,...,n
\label{1a}
\end{eqnarray}
where $W_n(z)\equiv P_n(z)-E$ and $z_i,\;i=1,...,n$, are roots of $W_n(z)$.
The roots $z_i,\;i=1,...,n$, are called also turning points (TP's).

By the above assumptions and definition the Stokes lines and the corresponding Stokes graph are defined on the
two sheeted Riemann surface $R_2$ with the turning points of $W_{n}(z)$ as the branch points of this surface.
However since on these two sheets the values of $\sqrt{W_{n}(z)}$ differ by signs only the projections on the
$z$-plane of the Stokes lines defined on each sheet coincide.

Therefore considering a pattern of SL's on the cut $z$-plane $C_{cut}$ with cuts emerging from the turning points of
$W_{n}(z)$ we see that the SL's on $C_{cut}$ are quasi continuous on the cuts despite
the fact that they are pieces of different SL's collected from the two sheets of $R_2$.

In general, a SL emerging from a given turning point $z_i$ can run to infinity of $C_{cut}$
or end at another turning point $z_j$. The SL with the first property is called {\it external} while with the second one
is called the {\it inner} SL.

A SG is called {\it critical} if at least one of its SL's is the inner one. It is called {\it
not critical} in the opposite case.

There are three SL's emerging from each TP at equal angles on $C_{cut}$.

In each SG there are at least $n+2$ SL's which are external, i.e. at least $n+2$ SL's have to escape to infinity of
$C_{cut}$ in $n+2$ different directions, i.e. there are $n+2$ asymptotes corresponding to these SL's. These asymptotes
which emerge from $z=0$ point of $C_{cut}$ at equal angles are called Stokes rays (SR).

There is of course the Euler-like relation between a number $e$ of the external SL's and
a number $i$ of the internal ones, namely $e+2i=3n$.

As it was said external SL's group into $n+2$ bunches. In each such a bunch with a given asymptote there is a finite
number of external SL's which can be enumerated with the clockwise rule. The first SL in each bunch and its last one are
the most
important. This is because when moving along the last SL from one bunch toward the closest TP which it emerges from and
avoiding clockwise all the TP met along such travelling to get successive SL's one finds the last SL to be linked
continuously with the first SL of the next bunch by a chain of internal SL's. These two SL's together with all these
internal ones passed by the travelling mentioned form a boundary of an infinite domain called a {\it sector}.

It follows from their definition that each sector does not contain any TP inside. There are therefore $n+2$ such sectors
for each SG on $C_{cut}$ separated by bunches. We can enumerate the sectors clockwise starting with a sector chosen
arbitrarily.

$C_{cut}$ has its partner $C_{cut}'$ with which it can be glued by cuts forming in this way the connected Riemann surface
$R_2$.

On $C_{cut}'$ the construction of sectors can be repeated so that on $R_2$ there are in this way $2n+4$ sectors which can
be enumerated clockwise in the way mentioned above.

However as we have mentioned earlier the sectors lying on $C_{cut}'$ are projections of the corresponding sectors lying on
$C_{cut}$ so that by a proper enumeration mentioned we can have the relations $S_{n+2+k}=PS_k,\;k=1,...,n+2$, where
$S_{n+2+k}$ and $S_k$ denote the sectors in $C_{cut}'$ and $C_{cut}$ respectively and $P$ denotes the projection.

By their definition in each sector $\Re\ll(\lambda \int_{z_i}^{z} \sqrt{W_n(\xi)}d\xi\r)$ has a definite sign ($\pm$) if
a root $z_i$ is chosen on its boundary. This sign alternates when $z$ is moving from a given sector to its neighbours or when it
crosses a cut. In particular
the corresponding signs are opposite for the pairs of sectors $S_k$ and $S_{n+2+k},\;k=1,...,n+2.$

There are $2n+4$ fundamental solutions (FS) to the equation \mref{1}.
In the sector $S_k,\;k=1,...,2n+4,$ the corresponding FS $\psi_k(z,\lambda)$ has the following form:
\be
\psi_k(z,\lambda) = W_n^{-\frac{1}{4}}(z)e^{\sigma_k\lambda {\tilde W}_n(z,z_k)}{\chi_k(z,\lambda)}
\label{10}
\end{eqnarray}
where $z\in S_k$ and $z_k$ is a turning point (a root) lying on the boundary of $S_k$ while
\begin{eqnarray}
{\tilde W}_n(z,z_k)=\int_{z_k}^z\sqrt{W_n(y)}dy\nn\\
\chi_{k}(z,\lambda) = 1 + \sum_{n{\geq}1}
\left( \frac{\sigma_k}{2\lambda} \right)^{n}Y_{k,n}(z,\lambda)
\label{11}
\ee
and
\be
 Y_{k,n}(z,\lambda) =\int_{\gamma_k(z)}d{y_{1}}
\int_{\gamma_k(y_1)}d{y_{2}} \ldots
\int_{\gamma_k(y_{n-1})}d{y_{n}}
\omega(y_{1})\omega(y_{2}) \ldots \omega(y_{n}){\times}\nn\\
\ll( 1 - e^{-2\sigma_k\lambda {\tilde W}_n(z,y_1)} \right)
\left(1 - e^{-2\sigma_k\lambda {\tilde W}_n(y_1,y_2)}\right)\cdots
\ll(1 - e^{-2\sigma_k\lambda {\tilde W}_n(y_{n-1},y_n)}\right)\nn\\n\geq 1
\label{12}
\ee
with
\be
\omega(z) ={\frac{5}{16}}{\frac{W_n^{\prime 2}(z)}{W_n^{\frac{5}{2}}(z)}}  -
{\frac{1}{4}}{\frac{W_n^{\prime\prime}(x)}{W_n^{\frac{3}{2}}(x)}}=
{W_n^{- \frac{1}{4}}(z)} \left( {W_n^{- \frac{1}{4}}(z)} \right)^{\prime{\prime}}
\label{13}
\ee

The signatures $\sigma_k=\pm 1$ present in the formulae \mref{10}-\mref{12} are defined in each particular sector $S_k$
in such a way to ensure the inequality $\Re(\sigma_k\lambda{\tilde W}_n(z,z_k))<0$ to be
satisfied in this sector.

The integration paths $\gamma_k(z)$ in \mref {12} which start from the infinities of the corresponding sectors are
canonical i.e. they are chosen in such a way to satisfy the inequality
$\Re(\sigma_k\lambda{\tilde W}_n(y_j,y_{j+1}))\geq 0,\;y_j,y_{j+1}\in\gamma_k(z),$ for each factor of
the integrand in \mref{12}.

It is easy to note the following identities:
\be
\psi_k(z,\lambda)\equiv\pm i\psi_{n+2+k}(Pz,\lambda),\;\;\;\;\;\;\;k=1,...,n+2,\;\;\;\; z\in C_{cut}
\label{13a}
\ee
which come out from the definitions \mref{10}-\mref{13} of FS's.

On the other hand in each pair $\psi_k(z,\lambda),\psi_j(z,\lambda),\;k,j=1,...,2n+4,\;j\neq n+2+k$ of FS's the solutions
are linear independent.

We can conclude therefore that we can limit our attention to the FS's defined on $C_{cut}$ only, considering the FS's
$\psi_k(z,\lambda)$ and $\psi_{n+2+k}(z,\lambda),\;k=1,...,n+2,$ as the two different representations of the same solution
$\psi_k(z,\lambda),\;k=1,...,n+2,$ defined on $C_{cut}$, i.e. if $\psi_k(z,\lambda)$ is continued analytically on $C_{cut}$
and $z$ crosses a cut of $C_{cut}$ during this continuation then $\psi_k(z,\lambda)$ has to be substituted after such
crossing by its alternative form $\psi_{n+2+k}(z,\lambda)$ satisfying \mref{13a}. This procedure has to be repeated each time when such
a crossing is happened.

Every point $z,\;z\in C_{cut}$, which can be achieved by a canonical path $\gamma_k(z)$ will be called canonical with
respect to the sector $S_k$ and the FS $\psi_k(z,\lambda)$. In particular TP's which belong to $\p S_k$ are canonical.

A domain $D_k$ of validity of the representation \mref{10}-\mref{13} of $\psi_k(z,\lambda)$ is called canonical. It means
that $D_k$ is collected of all canonical points of $\psi_k(z,\lambda)$ except the canonical TP's in which the representation
\mref{10}-\mref{13} of $\psi_k(z,\lambda)$ is singular.

Obviously a boundary $\p D_k$ is collected of TP's and SL's emerging from the former.

We call (after Eremenko {\it et al}
\cite{12}) all the SL's contained in $\p D_k$ the exceptional SL's if the corresponding SG is non-critical or if it is
critical with a single internal SL only.

Let us now characterize the ESL's for the cases mentioned.

Consider the non-critical SG.

The boundary $\p D_k$ of $D_k$ has to contain some number of TP's including the ones of $\p S_k$. Let
$z_r$ be such a TP. Then there is a canonical path $\gamma_k(z_r)$. Suppose that for any $z\in\gamma_k(z_r),\;z\neq z_r$,
$\Re(\sigma_k\lambda{\tilde W}_n(z_r,z))>0$. Then it follows that $\gamma_k(z_r)$ does not cross any SL's emerging from
$z_r$ and $\gamma_k(z_r)\in D_{L_{z_r}'L_{z_r}''}$, where $L_{z_r}'$ and $L_{z_r}''$ are two external SL's emerging from $z_r$ and
$D_{L_{z_r}'L_{z_r}''}$ is a domain of $C_{cut}$ with these two SL's as its boundary. However we can always deform
$\gamma_k(z_r)$ to make it coinciding partly with each of these two SL's up to $z_r$ and each of these common parts can be arbitrarily long.
We conclude therefore that these two SL's emerging from $z_r$ have to belong to $D_k$. It should be now clear that this is
the remaining external SL $L_{z_r}'''$ emerging from $z_r$ which is one of ESL's constructing $\p D_k$, i.e.
$L_{z_r}'''$ is this SL which points can not be achieved by any canonical path
$\gamma_k(z)$ if none of the SL's $L_{z_r}'$ and $L_{z_r}''$ is to be crossed by such paths.

For a given FS $\psi_k(z,\lambda)$ and
a given TP $z_r$ denote this exceptional SL by $L_k^r(\equiv L_{z_r}''')$. Then we have: $\p D_k=\bigcup_{r=1}^nL_k^r$.

Next consider the critical SG with a single internal SL between the TP's $z_{r_1}$ and $z_{r_2}$ and let $\psi_{k_1}(z,\lambda)$
be FS's defined in the respective sector $S_{k_1}$ with $z_{r_1}\in \p S_{k_1}$ while $z_{r_2}\in \p S_{k_2}$ where
$S_{k_2}$ is the sector where the FS $\psi_{k_2}(z,\lambda)$ is defined. Denote also by $L_{r_1r_2}$ the internal SL and by
$L_{r_j}'$ and $L_{r_j}''$ the remaining two external SL's emerging from the point $z_{r_j},\;j=1,2$. Then we have:
$\p D_k=\bigcup_{r=1}^nL_k^r,\;k\neq k_1,k_2$,
$\p D_{k_1}=\bigcup_{r=1,r\neq r_1,r_2}^nL_{k_1}^r\cup L_{r_1r_2}\cup L_{r_2}'\cup L_{r_2}''$ and
$\p D_{k_2}=\bigcup_{r=1,r\neq r_1,r_2}^nL_{k_2}^r\cup L_{r_1r_2}\cup L_{r_1}'\cup L_{r_1}''$.

Also in the considered critical case of SG we shall call each component of the sum constructing $\p D_k$ the
exceptional SL (ESL) corresponding to the solution $\psi_k(z,\lambda),\;k=1,...,n+2$.

For a given $\p D_k$ denote further by $V_k(\epsilon)$ its $\epsilon$-vicinity defined by the following
conditions:
\begin{enumerate}
\item $V_k(\epsilon)$ is a subset of $D_k$
\item $V_k(\epsilon)$ consists of all the points of $D_k$ the Euclidean distance of which to $\p D_k$ is smaller than $\epsilon$
(see Fig.2 and 3.)
\end{enumerate}

Let us still denote by $D_{k,\epsilon}$ a subset of $D_k$ given by $D_{k,\epsilon}=D_k\setminus{\bar V}_k(\epsilon)$.

The following {\bf Lemma} and {\bf Theorem 1} can be proved along the same lines as in \cite{11}:

\vskip 18pt

{\bf Lemma}

{\it In the domain $D_{k,\epsilon}$ the factor $\chi_k(z,\lambda)$ of the solution} \mref{10} {\it satisfies the following bound
\be
|\chi_k(z,\lambda)-1|\leq e^\frac{C_{\epsilon}}{\lambda_0}-1,\;\;\;\;\;|\lambda|>\lambda_0\nn\\
C_{\epsilon}=\liminf_{\gamma_k(z),\;z\in D_{k,\epsilon},\;k=1,...,n+2}\int_{\gamma_k(z)}|\omega(\xi)d\xi|<\infty
\label{14}
\ee
where $\gamma_k(z)$ are canonical}.

\vskip 30pt

{\bf Theorem 1}

{\it For sufficiently large $\lambda$ all zeros of $\psi_k(z,\lambda)$ lie entirely in the completion
$C_{cut}\setminus D_{k,\epsilon}$ of the domain $D_{k,\epsilon}$.}

Exactly in the same way as in \cite{11} {\bf Theorems 1a-1b} formulated below can be proved. However they need still
to consider asymptotic expansions of the $\chi$-factors of the fundamental solutions \mref{10} for $\lambda\to +\infty$
in their canonical domains. These asymptotic expansions are well defined and can be given the following
exponential forms (see \cite{11} as well as ref.1 of \cite{4} and ref.5 of \cite{5}):
\be
\chi_k(z,\lambda)\sim\chi_k^{as}(z,\lambda)=1 + \sum_{m{\geq}1}
\left( \frac{\sigma_k}{2\lambda} \right)^{n}{\tilde Y}_{k,m}(z)=
\exp\ll(\sum_{m{\geq}1}\left(\frac{\sigma_k}{2\lambda}\right)^m\int_{\infty_k}^zX_m(y)dy\r)
\label{551}
\ee
where
\be
{\tilde Y}_{k,m}(z)=\int_{\infty_k}^zdy_mW_n^{-\frac{1}{4}}(y_m)\ll(W_n^{-\frac{1}{4}}(y_m)
\int_{\infty_k}^{y_m}dy_{m-1}W_n^{-\frac{1}{4}}(y_{m-1})\times\r.\nn\\
\ll(...\ll.W_n^{-\frac{1}{4}}(y_2)\int_{\infty_k}^2dy_1W_n^{-\frac{1}{4}}(y_1)
\ll(W_n^{-\frac{1}{4}}(y_1)\r)''...\r)''\r)''
\label{552}
\ee
and
\be
\sum_{m{\geq}1}\left(\frac{\sigma_k}{2\lambda}\right)^mX_m(y)\equiv Z_k(y,\lambda)=
\frac{1}{\chi_k^{as}(z,\lambda)}\frac{d\chi_k^{as}(z,\lambda)}{dz}
\label{553}
\ee

Note that $X_m(y),\;m\geq 1$, are sector independent point functions on the
$C_{cut}$-plane being given by the following recurrent formula (see ref.2 of \cite{4}):
\be
X_1(z)=-\omega(z)=-W_n^{-\frac{1}{4}}(z)\ll(W_n^{-\frac{1}{4}}(z)\r)''=W_n^{-\fr}(z)\frac{U_{2n-2}(z)}{W_n^2(z)}\nn\\
X_m(z)=\fr W_n^{-\frac{3}{2}}(z)W_n'(z)X_{m-1}(z)-W_n^{-\fr}(z)X_{m-1}'(z)-\nn\\
W_n^{-\fr}(z)\sum_{k=1}^{m-2}X_kX_{m-k-1},\;\;\;
m=2,3,...
\label{554}
\ee
where $U_{2n-2}(z)$ is a polynomial of the $2n-2$-degree.

It follows from \mref{554} that $X_{2m},\;m\geq 1,$ have only poles at the turning points while $X_{2m+1},\;m\geq 0,$
have there the square root branch points. Therefore $Z^+(z,\lambda)$ and $Z^-(z,\lambda)$
where $Z^+(z,\lambda)+\sigma_kZ^-(z,\lambda)=Z_k(z,\lambda)$ have the same properties at these points respectively and
\be
Z^+(z,\lambda)=\sum_{m{\geq}1}\left(\frac{1}{2\lambda}\right)^{2m}X_{2m}(z)\nn\\
Z^-(z,\lambda)=\sum_{m{\geq}0}\left(\frac{1}{2\lambda}\right)^{2m+1}X_{2m+1}(z)
\label{555}
\ee

If we now take into account that $\chi_{i\to j}(\lambda)\equiv\lim_{z\to\infty_j}\chi_i(z,\lambda)=
\chi_{j\to i}(\lambda)$ where $\infty_j$ is the infinite point of the sector $S_j$ communicated canonically with the sector $S_i$
(see  ref.1 of \cite{4} and ref.5 of \cite{5}) then we get
\be
e^{\int_{\infty_i}^{\infty_j}(Z^++\sigma_iZ^-)dz}=e^{\int_{\infty_j}^{\infty_i}(Z^++\sigma_jZ^-)dz}
\label{556}
\ee

Since however $\sigma_i=-\sigma_j$ then we get from \mref{556}
\be
\int_{\infty_i}^{\infty_j}Z^+(z,\lambda)dz=0
\label{557}
\ee
for any pair of canonically communicated sectors.

However the integration in \mref{557} is now not limited by canonical paths since under the integral there are now no
exponentials limiting this integration to canonical paths, i.e. these paths can be freely deformed with the integral
beeing still
convergent. It is easy to see that because of that a given integral \mref{557} can be deformed to any other integral
between any pair of
sectors (i.e. not necessarily communicated canonically) as well as to any integral along an arbitrary loop. It means that
residua of $Z^+(z,\lambda)$ at the poles which it has at the turning points vanish.

Therefore we conclude that the Riemann surface of $Z^+(z,\lambda)$ is just the $C$-plane on which it is meromorphic with
vanishing residua at its poles. It means of course that this is the property of each $X_{2m},\;m\geq 1$ as well. Thus
when integrating $Z_k(z,\lambda)$ along contours starting and ending at the same points we get only contribution from
the odd part of $Z_k(z,\lambda)$, i.e. from $\sigma_kZ^-(z,\lambda)$.

We can now formulate two theorems being analogous with the corresponding {\bf Theorems 3a-3b} of \cite{11}.

{\bf Theorem 1a}

{\it Zeros $\zeta_{l,qr}^{(k)}(\lambda)$ of $\psi_k(z,\lambda)$, $\lambda=|\lambda|e^{i\beta},\;|\lambda|=[|\lambda|]+\Lambda,\;0\leq\Lambda<1,\;k,l=1,...,n+2,\;q=0,1,2,...,\;
r=0,1,...,[|\lambda|]-1$, in the non-critical cases and in the regular limit $\lambda\to\infty$ (with constant value of
$\Lambda$), are distributed on $C_{cut}$ uniquely along the corresponding exceptional SL's according to the formulae}:
\be
\int_{K_l(\zeta_{l,qr}^{(k)}(\lambda))}\ll(\fr\sqrt{W_n(y)}+
\frac{1}{2\lambda}Z^-(y,\lambda)\r)dy=\sigma_k\ll(q[|\lambda|]+r-\frac{1}{4}\r)\frac{i\pi}{\lambda}
\label{14a}
\ee
{\it where $K_l(\zeta_{l,qr}^{(k)}(\lambda))$ is a contour which starts and ends at
$\zeta_{l,qr}^{(k)}(\lambda)$ rounding the turning point $z_l$ anticlockwise (see} Fig.1{\it).

Zeros $\zeta_{l,qr}^{(k)}(\lambda)$ for $q>0$ have the following semiclassical expansion (see Appendix A):}
\be
\zeta_{l,qr}^{(k)}(\lambda)=\sum_{p\geq 0}\frac{1}{\lambda^p}\zeta_{l,qrp}^{(k)}(\Lambda)
\label{141}
\ee
{\it with the two first terms given by}:

\vskip 18pt

\begin{tabular}{c}
\psfig{figure=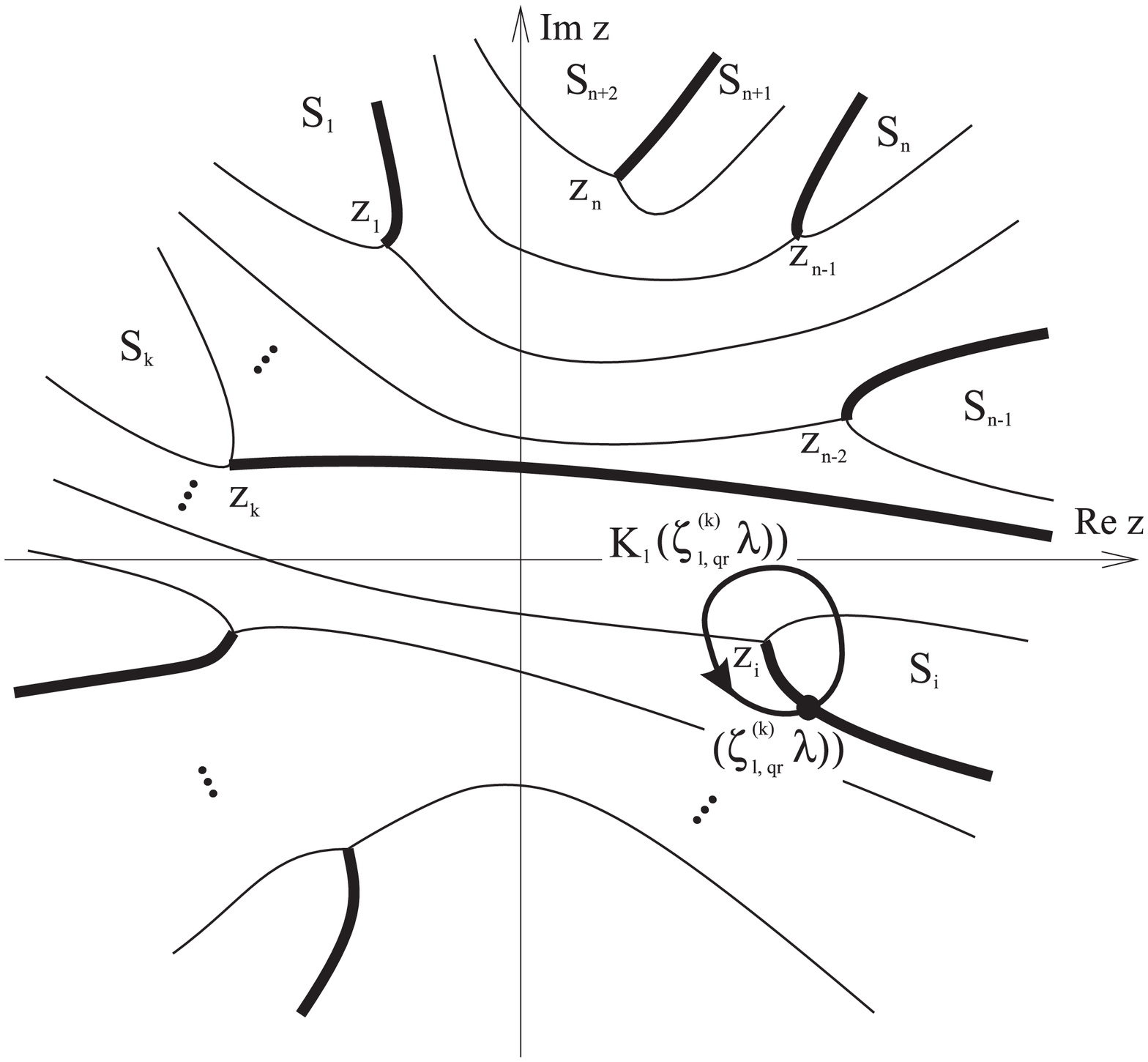,width=11cm}\\
Fig.1  Ecxeptional lines (bold Stokes lines), zeros $\zeta_{l,qr}^{(k)}(\lambda)$ of $\psi_k(z,\lambda)$ and \\the itegration
contour $K_l(\zeta_{l,qr}^{(k)}(\lambda))$
\end{tabular}

\vskip 18pt

\be
\int_{K_l(\zeta_{l,qr0}^{(k)})}\fr\sqrt{W_n(y)}dy=
\int_{z_l}^{\zeta_{l,qr0}^{(k)}}\sqrt{W_n(y)}dy=\sigma_kqi\pi e^{-i\beta}\nn\\
\zeta_{l,qr1}^{(k)}(\Lambda)=\sigma_k(r-q\Lambda-\frac{1}{4})\frac{i\pi e^{-i\beta}}{\sqrt{W_n(\zeta_{l,qr0}^{(k)})}}
\label{142}
\ee

{\it For $q=0$ we have instead $\zeta_{l,0r0}^{(k)}(\Lambda)\equiv z_l$ and}
\be
\int_{z_l}^{z_l+\zeta_{l,0r1}^{(k)}/\lambda}\sqrt{W_n(y)}dy=
(r-\frac{1}{4})\frac{i\pi}{\lambda}\nn\\
r>\frac{|\lambda|}{\pi}\limsup_{|\phi|\leq\pi}\ll|\int_{z_l}^{z_l+\epsilon e^{i\phi}}\sqrt{W_n(y)}dy\r|
\label{142b}
\ee
{\it as well as}
\be
\zeta_{l,0r2}^{(k)}=\frac{1}{8}\frac{\int_{K_l(z_l+\zeta_{l,0r1}^{(k)}/\lambda)}X_1(y)dy}
{\sqrt{W_n(z_l+\zeta_{l,0r1}^{(k)}/\lambda)}}
\label{142c}
\ee
{\it with $|\zeta_{l,0r1}^{(k)}/\lambda|>\epsilon$}.

Assuming now that there is a single inner SL between the roots $z_{r_1}\in\p S_{k_1}$ and $z_{r_2}\in\p S_{k_2}$ while all
the remaining SL's of the SG corresponding to $W_n(z)$ are external and putting:
\be
\sigma_{k_1}\lambda_s(R)\int_{z_{r_1}}^{z_{r_2}}\sqrt{W_n(y)}dy\equiv
\sigma_{k_1}\lambda_s(R)I_{r_1r_2}=-\fr(s+R)i\pi\nn\\s=0,1,2,...,\;-\fr\leq |R|<\fr
\label{24}
\ee
we get the following theorem analogous with {\bf Theorem 3b} of \cite{11}:

\vskip 18pt

{\bf Theorem 1b}

{\it Zeros $\zeta_{l,qr}^{(k)}(\lambda)$ of $\psi_k(z,\lambda)$, $k=1,...,n+2,\;l=1,...,n,\;q=1,2,...,\;
r=0,1,...,[|\lambda|]-1$ in the critical cases and in the regular limit $\lambda\to\infty$ are distributed on $C_{cut}$
uniquely along the corresponding exceptional SL's according to the formulae:

a) $(k,l)\neq (k_1,r_1),(k_1,r_2),(k_2,r_1),(k_2,r_2)$, $|\lambda|=[|\lambda|]+\Lambda,\;0\leq\Lambda<1$ and $\Lambda$ is fixed}
\be
\int_{K_l(\zeta_{l,qr}^{(k)}(\lambda))}\ll(\fr\sqrt{W_n(y)}+
\frac{1}{2\lambda}Z^-(y,\lambda)\r)dy=\sigma_k\ll(q[|\lambda|]+r-\frac{1}{4}\r)\frac{i\pi}{\lambda}
\label{14b}
\ee
{\it where $K_l(\zeta_{l,qr}^{(k)}(\lambda))$ is a contour which starts and ends at
$\zeta_{l,qr}^{(k)}(\lambda)$ rounding the turning point $z_l$ anticlockwise.

Zeros $\zeta_{l,qr}^{(k)}(\lambda)$ have the following semiclassical expansion:}
\be
\zeta_{l,qr}^{(k)}(\lambda)=\sum_{p\geq 0}\frac{1}{\lambda^p}\zeta_{l,qrp}^{(k)}(\Lambda)
\label{143}
\ee
{\it with two first terms given by}:
\be
\int_{K_l(\zeta_{l,qr0}^{(k)})}\fr\sqrt{W_n(y)}dy=
\int_{z_l}^{\zeta_{l,qr0}^{(k)}}\sqrt{W_n(y)}dy=\sigma_kqi\pi e^{-i\beta}\nn\\
\zeta_{l,qr1}^{(k)}(\Lambda)=
\sigma_k(r-q\Lambda-\frac{1}{4})\frac{i\pi e^{-i\beta}}{\sqrt{W_n(\zeta_{l,qr0}^{(k)})}}
\label{144}
\ee

{\it For $q=0$ we have instead $\zeta_{l,0r0}^{(k)}(\Lambda)\equiv z_l$ and}
\be
\int_{z_l}^{z_l+\zeta_{l,0r1}^{(k)}(\Lambda)/\lambda}\sqrt{W_n(y)}dy=
(r-\frac{1}{4})\frac{i\pi}{\lambda}\nn\\
r>\frac{|\lambda|}{\pi}\limsup_{|\phi|\leq\pi}\ll|\int_{z_l}^{z_l+\epsilon e^{i\phi}}\sqrt{W_n(y)}dy\r|
\label{144a}
\ee
{\it as well as}
\be
\zeta_{l,0r2}^{(0)}(\Lambda)=\frac{1}{8}\frac{\int_{K_l(z_l+\zeta_{l,0r1}^{(k)}(\Lambda)/\lambda)}X_1(y)dy}
{\sqrt{W_n(z_l+\zeta_{l,0r1}^{(k)}(\Lambda)/\lambda)}}
\label{144b}
\ee

{\it b) $(k,l)=(k_1,r_1),(k_1,r_2),(k_2,r_1),(k_2,r_2),\; |\lambda_s(R)|=[|\lambda_s(R)|]+\Lambda_s(R),\;
0\leq\Lambda_s(R)<1,\;s=0,1,2,...,$ and $R$ fixed where $\lambda_s(R)=|\lambda_s(R)|e^{i\beta_s(R)}=-\frac{s+R}{I_{r_1r_2}}i\pi\sigma_k$
($\Lambda_s(R)$ is bounded but not fixed)

In these cases the number
$q$ is bounded, i.e. $q\leq |I_{r_1r_2}|/\pi$ and in} \mref{144} {\it the minus sign is to be chosen according to our earlier
conventions.

In the regular limit $\lambda_s(R)\to\infty$ there are
two infinite sequences of zeros $\zeta_{r_2,qr}^{(k_1)\pm},\;q=1,2,...,\;r=0,1,...,[|\lambda_s(R)|]-1$, of
$\psi_{k_1}(z,\lambda_s(R))$ distributed along the two ESL's $L^{r_2'}$ and $L^{r_2''}$ of the sector $S_{k_2}$
according to the following rules}:
\be
\int_{K_{r_2}(\zeta_{r_2,qr}^{(k_1)\pm})}\ll(\fr\sqrt{W_n(y)}+\frac{1}{2\lambda_s(R)}Z^-\r)dy
=\pm\sigma_{k_1}\ll(q[|\lambda_s(R)|]+r-\frac{1}{4}+\frac{R}{2}\r)\frac{i\pi}{\lambda_s(R)}+\nn\\
\pm\frac{1}{4\lambda_s(R)}\oint_{K_{r_1r_2}}Z^-dy-
\frac{\sigma_{k_1}}{2\lambda_s(R)}\ln2\cos\ll(R\pi-\frac{i\sigma_{k_1}}{2}\oint_{K_{r_1r_2}}Z^-dy\r)
\label{14c}
\ee
{\it with the closed contour $K_{r_1r_2}$ surrounding the internal SL $L_{r_1r_2}$ and with the following first
coefficients of the corresponding semiclassical expansion of $\zeta_{r_2,qr}^{(k_1)\pm}$}:
\be
\int_{z_{r_2}}^{\zeta_{r_2,qr0}^{(k_1)\pm}}\sqrt{W_n(y)}dy=\mp\sigma_{k_1}qi\pi e^{-i\beta_s(R)}\nn\\
\zeta_{r_2,qr1}^{(k_1)\pm}(R)=\mp\sigma_{k_1}\ll(r-q\Lambda_s(R)-\frac{1}{4}+\frac{R}{2}
\pm\fr\ln2\cos(R\pi)\r)\frac{i\pi e^{-i\beta_s(R)}}{\sqrt{W_n(\zeta_{r_2,qr0}^{(k_1)\pm})}}
\label{145}
\ee
{\it where the plus sign corresponds to a vicinity of the SL being the upper boundary of $S_{k_2}$ while the minus one
to a vicinity of its lower boundary.

Mutatis mutandis the same results are valid for the solution $\psi_{k_2}(z,\lambda_s(R))$}

Fig.2 shows all ESL's occupied by zeros of $\psi_{k_1}(z,\lambda_s(R))$ in the limits considered in the above theorem.

\section{The quantized case of the limit $\hbar\to 0$}

\hskip+2em Let us consider now the $\hbar$-quantized case. We would like to consider this limit a little bit more
generally than it was done by Hezari \cite{9} assuming that the
quantization condition is provided by identifying the fundamental solutions defined in the sectors $S_{k_1},S_{k_2}$.
Such an assumption needs however a comment since it can appear that the corresponding quantization condition does not
exist at all in the limit $\lambda(=2m\hbar^{-1})\to\infty$.

First let us note that if the polynomial potential is fixed and $W_n(z)$ has only simple roots then the argument of
$\hbar$ remains as the unique tool for changing a form of the corresponding SG, i.e. this SG is independent of the
absolute value of $\hbar$. On the other hand changing $\arg\hbar$ one can find only a finite number of its values for which
the corresponding SG is critical, i.e. for all except a final number of values of $\arg\hbar$ these SG's are
non-critical. But for the non-critical SG's each two sectors of such graphs communicate canonically, what means that
each FS can be continued to each sector directly along canonical path. This means further however that to vanish in any
other sector except the one the FS considered is defined in it is necessary for its factor $\chi(z,\lambda)$ to vanish
for $z\to\infty$ in such a sector. But since $\chi(z,\lambda)\to 1$ along any canonical path for $|\lambda|\to\infty$
the latter vanishing can not happen for $|\lambda|$ sufficiently large, i.e. for such $\lambda$ there are no solutions
for the $\lambda$-eigenvalue problem when SG's are non-critical.

Therefore we have to conclude that solutions of the $\lambda$-eigenvalue problem with
arbitrarily large eigenvalues of $\lambda$ can exist only for critical cases of SG's and two sectors can not
communicate canonically if FS's which are defined in them are to coincide.

\vskip 15pt

\begin{tabular}{c}
\psfig{figure=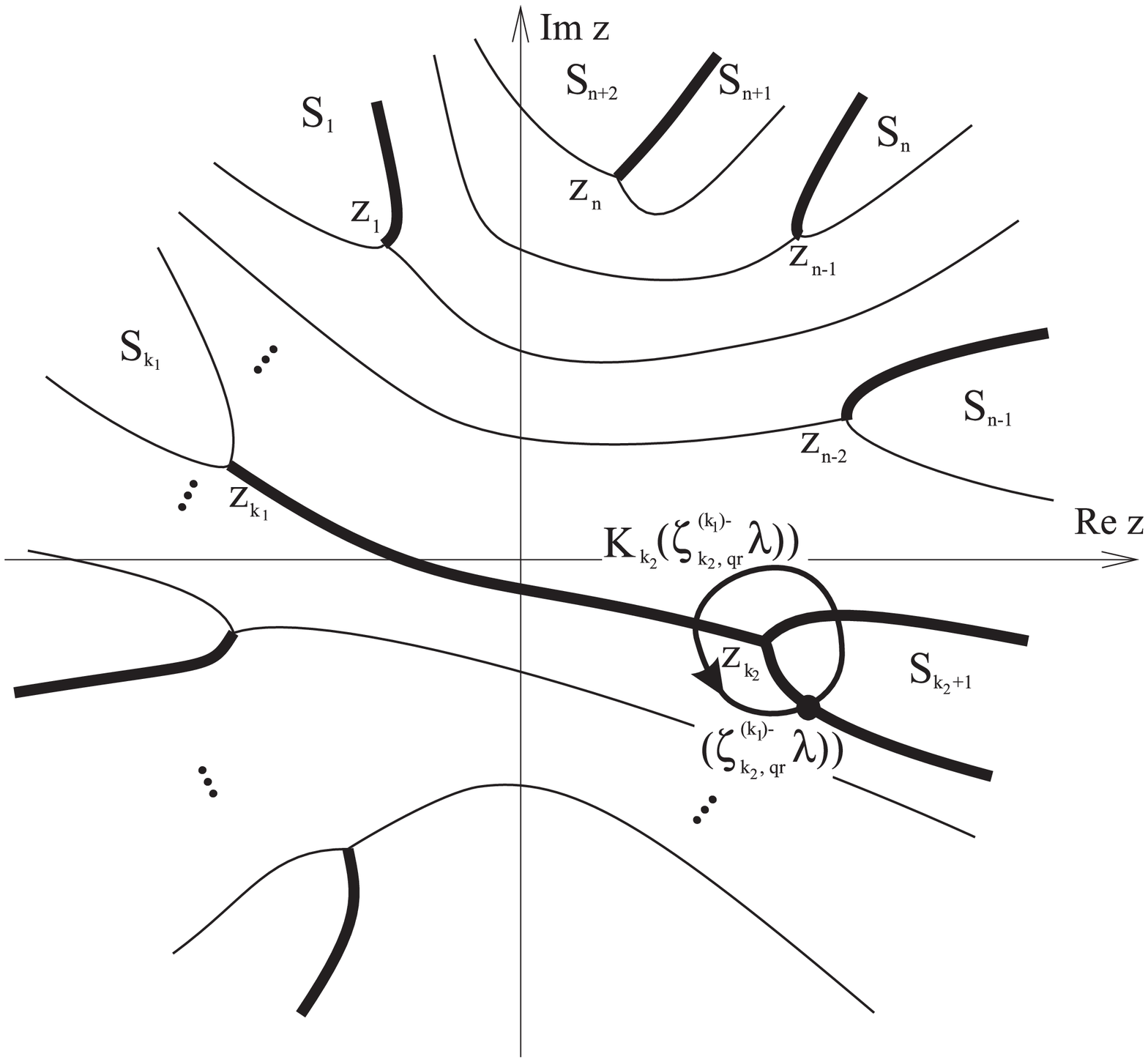,width=11cm}\\
Fig.2  Ecxeptional lines (bold Stokes lines), zeros $\zeta_{k_2,qr}^{(k_1)-}(\lambda_s(R))$ of $\psi_{k_1}(z,\lambda_s(R))$\\ and
the integration contour $K_{k_2}(\zeta_{k_2,qr}^{(k_1)-}(\lambda_s(R)))$ in the non-quantized critical case
\end{tabular}

\vskip 15pt

However as we have already mentioned above there is only a finite number of $\arg\lambda$ for which SG's can be critical.
For a given such $\arg\lambda$ (and fixed polynomial) the only variable which can be quantized then is $|\lambda|$, i.e.
a real quantity. But a quantization condition which one gets by matching two FS's is in general a complex number
equation, which means that one gets typically two real conditions for one real variable. Such conditions can not be
satisfied in general by one real variable only. Also a change of $\arg\lambda$ from its one discret value to another seems
not to save the situation.

Therefore one needs some additional assumptions which allow one to eliminate somehow one
of these two real conditions. As such assumptions can be chosen for example the realness of polynomial potentials used
together with matching two real FS's which are allowed by the real polynomial potentials. One can convince oneself that
in such a case the corresponding quantization equation is a real one (see for example \mref{52}).

Consequently to proceed further we have to assume that we have introduced proper conditions for a polynomial considered
as well as the choice of solutions made above has been proper. Then it is not difficult to get the following asymptotic
quantization condition for FS's defined in the sectors $S_{k_1},S_{k_2}$:
\be
1+\exp\ll[\sigma_{k_1}\oint_{K_{r_1r_2}}\ll(\lambda\sqrt{W_n(y)}+Z^-(y,\lambda)\r)dy\r]=0
\label{25}
\ee
or
\be
\oint_{K_{r_1r_2}}\ll(\lambda_s\sqrt{W_n(y)}+Z^-(y,\lambda_s)\r)dy=-\sigma_{k_1}(2s+1)i\pi,\;\;\;\;s=0,1,2,...
\label{26}
\ee

Compairing the last result with \mref{24} we see that $R=\fr$ in this formula and we can repeat arguments  of the
previous paper \cite{11} leading us to {\bf Corollary 1b} of this paper almost with no changes.

\vskip 20pt

{\bf Corollary 1}

{\it In the singular limit $\lambda_s\to\infty$ roots of FS's for the potential $P_n(z)$ with a single inner
SL which is quantized (in the sense of eq.}\mref{26}){\it are distributed
uniquely on the exceptional lines for both the quantized and not quantized solutions}

\section{The critical case -- a double-well potential. The not quantized case}

\hskip+2em In this section we would like to extend the results of the previous two sections to the cases of SG's with
two and more
internal SL's. However as it follows from the discussion accompanied the case of SG's with a single internal SL such an
extension can not be expected to be direct so that we shall limit initially ourselves to the cases of double-well (D-W)
real polynomial potentials and in particular to a real polynomial of tenth degree $P_{10}(z)$
since as we shall see for this case we can arrange SG shown in Fig.3 to be composed of all essential ingredients of a
general case.

However even simplifying the cosidered cases of potentials in the way described above we are still
faced with the problem of ESL's for these cases. In fact we can not define them as previously as forming a boundary of
the canonical
domain corresponding to a chosen FS since most of SL's which are the limit loci of zeros of this chosen FS
can lie outside this domain. Therefore to identify such SL's
as hypothetical ESL's one has to continue analytically the solution chosen outside its canonical domain in a way allowing one to take the
semiclassical limit in any stage of this continuation. The latter demand means that such analytical continuations have
to be done along canonical paths. This is always possible when all turning points are simple. In many of our earlier
papers we have shown how such analytical continuations can be done with the help of other FS's (see, for example, ref.4
of \cite{4} or ref.1,2 of \cite{5} for the respective procedure). It has to be stressed also that each FS given in the
form \mref{10}-\mref{13} can be continued
analytically in this way to any point of $C_{cut}$ except the TP's in which the form \mref{10}-\mref{13} is singular.

\vskip 18pt

\begin{tabular}{c}
\psfig{figure=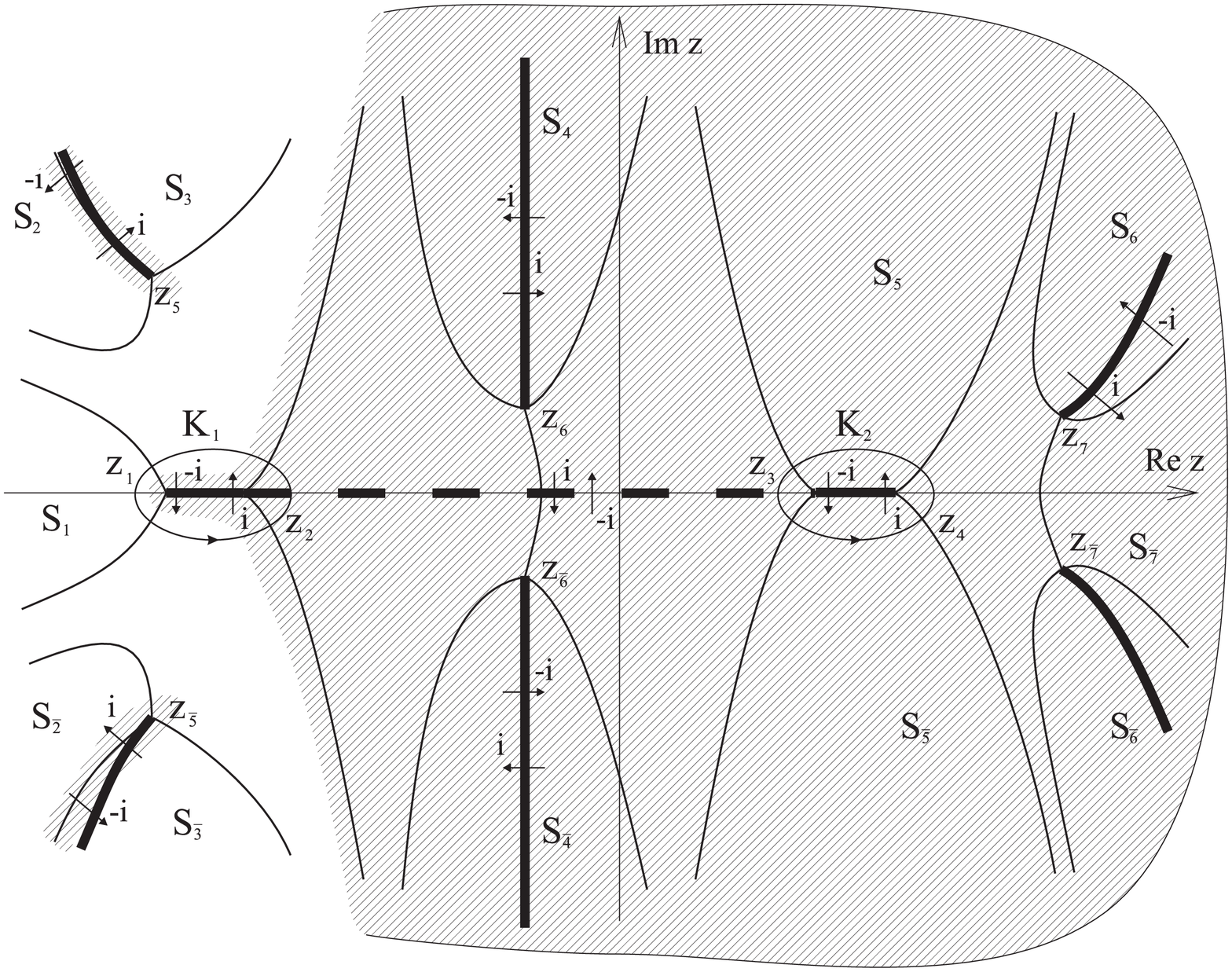,width=11cm}\\
Fig.3  The Stokes graph for the double-well potential $P_{10}(z)$. The bold full lines\\
denote cuts for $W_{10}^{\fr}(z)$
and $W_{10}^{-\frac{1}{4}}(z)$ while the bold dashed line - for $W_{10}^{-\frac{1}{4}}(z)$.\\
The outlined domains which are not communicating canonically with \\the sector $S_1$ contain possible
ESL's of $\psi_1(z,\lambda)$.
\end{tabular}

\vskip 18pt

On the other hand since we are going to compare configurations of loci of two chosen FS's in the cases when they are
linearly independent and when they are matched (quantizing in this way the Planck constant) then we see invoking the
discussion of the previous section that we have to choose such a pair of solutions properly to ensure this matching.

It seems therefore that in the case of the chosen potential $P_{10}(z)$ two obvious candidates of FS's for acheaving
this goal are the solutions $\psi_1(z,\lambda)$ and $\psi_7(z,\lambda)$ defined in the respective sectors $S_1$ and
$S_7$ of Fig.3. Below we shall investigate loci of zeros of these solutions and we shall call as previously exceptional
the lines along which these zeros are distributed and if they are SL's we shall maintain the previous description as
ESL's.

A general observation suggesting where the limit loci of zeros should be looked for in the critical cases of SG's is that
the latter arise in the case of a fixed polynomial potential from
non-critical ones by changing $\arg\lambda$ so that three SL's emerging from each root of the polynomial rotate around
the root. During these rotations different particular SL's coincide leading to critical SG's.

Consider the critical SG of Fig.3. It has appeared for $\arg\lambda=0$. However for sufficiently small $\epsilon$ and
$0<|\arg\lambda|<\epsilon$ all SG's corresponding to these $\lambda$'s are non-critical. For these SG's
{\bf Theorem 1a} applies so that for each $\psi_k(z,\lambda)$ its limit zeros distribution when $\lambda\to\infty$
coincides with its ESL's. If now we come back (by a rotation of SL's) to the critical configuration for
$\arg\lambda=0$ then some of these ESL's will coincide partly with some other SL's. All these
coinciding non-exceptional SL's become then the limit loci of zeros of the considered $\psi_k(z,\lambda)$ on these their
parts which coincide with the ESL's. {\bf Theorem 1b} is the illustration of this rule.

Applying the above suggestions to the SG of Fig.3 we readily get the corresponding limit distributions of zeros of
$\psi_1(z,\lambda)$ and $\psi_7(z,\lambda)$ not matched with themselves (i.e. for not quantized $\lambda$) in the forms
shown in Fig.4a and Fig.4b respectively.

Nevertheless while these suggestions are very convincing they have to be confirmed by detailed calculations.

Therefore it follows that we should continue analytically in the way mentioned above the solution $\psi_1(z,\lambda)$ ($\lambda$ is
now real positive) to the vicinities of all these SL's
emerging from the turning points $z_1,...,z_{\bar 7}$ with which the sector $S_1$ cannot communicate canonically,
looking for positions of possible zeros of this solution in the distinguished domains when the limit $\lambda\to+\infty$
is taken. We shall consider these positions for both the cases - the qunatized and the not quantized $\lambda$.

The relevant vicinities are
shown in Fig.3 as the outlined domains containing the corresponding SL's. However, because of the obvious symmetry of the
SG considered with respect to the real axis we can limit our investigations
only to the upper part of these domains including the SL's emerging from the turning points $z_1,...,z_7$
(i.e. with no bars over indeces).

As previously the main tool of making these investigations is the analytical continuation of the solution $\psi_1$ to the
respective domains along canonical paths expressing the solution by linear combinations of FS's which can contact with
these domains canonically. In our case this corresponds to make linear combinations of $\psi_1$ by the following pairs
of FS's: 1. ($\psi_2,\psi_3$), 2.($\psi_2,\psi_{\bar 2}$), 3.($\psi_3,\psi_4$), 4.($\psi_3,\psi_5$), 5.($\psi_4,\psi_5$),
6.($\psi_5,\psi_{\bar 5}$), 7.($\psi_5,\psi_6$) and 8.($\psi_6,\psi_7$).

Note that signatures of two FS's in each chosen
pair are different. This is quite important condition since the choice of two FS's with the same signatures would provide us
with some identities in the limit $\lambda\to\infty$ rather than with appropriate conditions for the limit loci of zeros
(see App.C).

We shall consider all the cases step by step.

In this section $\lambda$ is assumed to be real and positive and the limits $\lambda\to +\infty$ which we are going to
consider are regular only. These regular limits
are defined by sequences $\lambda_{s_i},\;s_i=0,1,2,...,\;i=1,2,$ of $\lambda$'s satisfying the conditions:
\be
\lambda_{s_1}\int_{z_1}^{z_2}\sqrt{W_{10}(y)}dy=-(s_1+R_1)i\pi,\;\;\;|R_1|<\fr,\;\;\;s_1=0,1,2,...
\label{261}
\ee
or
\be
\lambda_{s_2}\int_{z_3}^{z_4}\sqrt{W_{10}(y)}dy=(s_2+R_2)i\pi,\;\;\;|R_2|<\fr,\;\;\;s_2=0,1,2,...
\label{262}
\ee
for {\it fixed} $R_1$ or $R_2$ respectively or by the representation $\lambda=[\lambda]+\Lambda,\;0\leq\Lambda<1$ where
$[\lambda]$ is an integer part of $\lambda$ and $\Lambda$ is fixed.

The integrals in \mref{261}-\mref{262} are taken above the respective cuts.

To avoid possible inconsistences or errors in taking properly the limit $\lambda\to +\infty$ we first establish the
exact forms of the formulae providing us with conditions for loci of zeros of $\psi_1(z,\lambda)$. Next these exact conditions
are substituted by their full asymptotic expansions i.e. up to all orders in $\lambda^{-1}$ from which one could
select the lowest order of the semiclassical asymptotic in a way similar to that used in {\bf Theorem a,b}.

{\it The case 1.}

Taking into account the representation \mref{10} and the conventions accompanying it as well as Fig.1 we get
(see App.B or, for example, ref.4 of \cite{4} or ref.1,2 of \cite{5} for the respective procedure):
\be
\psi_1(z,\lambda)=\alpha_{\frac{1}{2}\to 3}\psi_2(z,\lambda)+\alpha_{\frac{1}{3}\to 2}\psi_3(z,\lambda)=\nn\\
         W_{10}^{-\frac{1}{4}}(z)e^{\lambda\int_{z_1}^{z_5}\sqrt{W_{10}(y)}dy}\ll(\chi_{1\to 3}(\lambda)\chi_2(z,\lambda)
          e^{\lambda\int_{z_5}^z\sqrt{W_{10}(y)}dy}+i\chi_3(z,\lambda)e^{-\lambda\int_{z_5}^{z}\sqrt{W_{10}(y)}dy}\r)
\label{271}
\ee
where the last representation of $\psi_1(z,\lambda)$ has been written in the sector $S_3$-side of the cut beginning at $z=z_5$ in
Fig.3. Here
$\alpha_{\frac{i}{j}\to{k}}=\lim_{z\to\infty_k}\frac{\psi_i(z,\lambda)}{\psi_j(z,\lambda)}$ and
$\chi_{i\to j}(\lambda)=\lim_{z\to\infty_j}\chi_i(z,\lambda)=\chi_{j\to i}(\lambda)$ where $\infty_{i,j}\in S_{i,j}$
respectively and the limits are calculated along canonical paths. For particularities of getting the eq.\mref{271} as well
as the others below see Appendix B.

Therefore for a distribution of zeros $\zeta_{5,m}^{(1)}(\lambda)$ of $\psi_1(z,\lambda)$ in the considered domain we get the condition:
\be
\int_{z_5}^{\zeta_{5,m}^{(1)}(\lambda)}\sqrt{W_{10}(y)}dy=(m-\frac{1}{4})\frac{i\pi}{\lambda}-\frac{1}{2\lambda}
\ln\frac{\chi_3(\zeta_{5,m}^{(1)}(\lambda),\lambda)}{\chi_{1\to 3}(\lambda)\chi_2(\zeta_{5,m}^{(1)}(\lambda),\lambda)}
\label{272}
\ee
with $m$ - an integer.

Taking now in \mref{272} the regular limit $\lambda(=[\lambda]+\Lambda)\to+\infty$ with fixed $\Lambda$ and noticing by \mref{551}-\mref{553} that asymptotically:
\be
\ll(\frac{\chi_3(\zeta_{5,m}^{(1)}(\lambda),\lambda)}{\chi_{1\to 3}(\lambda)\chi_2(\zeta_{5,m}^{(1)}(\lambda),\lambda)}\r)^{as}=
e^{-\int_{K_5{(\zeta_{5,m}^{(1)}}(\lambda))}Z^-(y,\lambda)dy}
\label{2721}
\ee
where $K_5{(\zeta_{5,m}^{(1)}}(\lambda))$ is a contour shown in Fig.3 which starts end ends at the point
$\zeta_{5,m}^{(1)}(\lambda)$ rounding the point $z_5$ anticlockwise
(this contour is not closed since it starts and finishes on different sheets of the corresponding Riemann surface) we
get (putting $m=q[\lambda]+r$):
\be
\int_{K_5{(\zeta_{5,qr}^{(1)}}(\lambda))}\ll(\fr\sqrt{W_{10}(y)}dy+\frac{1}{2\lambda}
Z^-(y,\lambda)dy\r)=(q[\lambda]+r-\frac{1}{4})\frac{i\pi}{\lambda}\nn\\q=0,1,2,3,...,\;r=0,1,...,[\lambda]-1
\label{2722}
\ee

The last formula is the exact implicite condition for a semiclassical asymptotic expansion of $\zeta_{5,m}^{(1)}(\lambda)$ in
$\lambda$ which therefore have the forms \mref{141}-\mref{142c} given in {\bf Theorem 1}.

{\it The case 2.}

The respective linear combination is:
\be
\psi_1(z,\lambda)=\alpha_{\frac{1}{2}\to{\bar 2}}\psi_2(z,\lambda)+\alpha_{\frac{1}{\bar 2}\to{2}}\psi_{\bar 2}(z,\lambda)=\nn\\
           \frac{W_{10}^{-\frac{1}{4}}(z)}{\chi_{2\to{\bar 2}}(\lambda)}
           \ll(-i\chi_2(z,\lambda)e^{-\lambda\int_{z_1}^z\sqrt{W_{10}(y)}dy}+
           \chi_{\bar 2}(z,\lambda)e^{\lambda\int_{z_1}^{z}\sqrt{W_{10}(y)}dy}\r)
\label{273}
\ee
where the last equation has been written above the cut between $z_1$ and $z_2$.

The asymptotic distribution of zeros $\zeta_{1,qr}^{(1)}(\lambda)$ in the vicinity of the inner SL between the points $z_1$ and $z_2$
when the regular limit $\lambda(=[\lambda]+\Lambda)\to+\infty,\;\Lambda$ - fixed, is taken is then given by:
\be
\int_{K_1{(\zeta_{1,qr}^{(1)}}(\lambda))}\ll(\fr\sqrt{W_{10}(y)}dy+\frac{1}{2\lambda}
Z^-(y,\lambda)dy\r)=(q[\lambda]+r-\frac{1}{4})\frac{i\pi}{\lambda}\nn\\q=0,1,2,...,\;r=0,1,...,[\lambda]-1
\label{274}
\ee

The number $q$ however is bounded by a
"length" $\int_{z_1}^{z_2}\sqrt{W_{10}(y)}dy=-(q_1+r_1)i\pi$, measured above the cut $z_1,z_2$ of the SL, with
integer $q_1\geq 0$ and $0\leq r_1<1$, i.e. $q\leq q_1$.

{\it The case 3.}

We get for this case:
\be
\psi_1(z,\lambda)=\alpha_{\frac{1}{3}\to{\bar 3}}\psi_3(z,\lambda)+\alpha_{\frac{1}{\bar 3}\to{3}}\psi_{\bar 3}(z,\lambda)=\nn\\
\ll(\alpha_{\frac{1}{3}\to{\bar 3}}+\alpha_{\frac{1}{\bar 3}\to{3}}\alpha_{\frac{\bar 3}{3}\to{4}}\r)\psi_3(z,\lambda)+
\alpha_{\frac{1}{\bar 3}\to{3}}\alpha_{\frac{\bar 3}{4}\to{3}}\psi_4(z,\lambda)
\label{34}
\ee
or
\be
\psi_1(z,\lambda)=-i\frac{e^{-\lambda\ll(\int_{z_2}^{z_6}-\fr\oint_{K_1}\r)\sqrt{W_{10}(y)}dy}}{\chi_{3\to{\bar 3}}(\lambda)}\times\nn\\
\ll(\chi_{1\to{\bar 3}}(\lambda)+
e^{-\lambda\oint_{K_1}\sqrt{W_{10}(y)}dy}\chi_{1\to 3}\chi_{{\bar 3}\to{4}}(\lambda)\r)\psi_3(z,\lambda)+\nn\\
e^{\lambda\ll(\int_{z_2}^{z_6}-\fr\oint_{K_1}\r)\sqrt{W_{10}(y)}dy}\chi_{{1}\to{3}}(\lambda)\psi_4(z,\lambda)=\nn\\
\frac{W_{10}^{-\frac{1}{4}}(z)}{\chi_{3\to{\bar 3}}(\lambda)}\ll[-iE_1
e^{\lambda\ll(\fr\oint_{K_1}-\int_{z_2}^{z}\r)\sqrt{W_{10}(y)}dy}\chi_3(z,\lambda)+\r.\nn\\
\ll.\chi_{{1}\to{3}}(\lambda)\chi_{3\to {\bar 3}}(\lambda)
e^{-\lambda\ll(\fr\oint_{K_1}-\int_{z_2}^z\r)\sqrt{W_{10}(y)}dy}\chi_4(z,\lambda)\r]
\label{35}
\ee
where
\be
E_1=\chi_{1\to{\bar 3}}(\lambda)+e^{-\lambda\oint_{K_1}\sqrt{W_{10}(y)}dy}\chi_{1\to 3}(\lambda)\chi_{{\bar 3}\to{4}}(\lambda)
\label{351}
\ee
It follows from the formula \mref{35} that zeros $\zeta_2^{(1)}(\lambda)$ of $\psi_1(z,\lambda)$ in the outlined domain of Fig.4
containing the sector $S_4$ have to satisfy the equation:
\be
e^{2\lambda\int_{z_2}^{\zeta_{2}^{(1)}(\lambda)}\sqrt{W_{10}(y)}dy}=
ie^{\lambda\oint_{K_1}\sqrt{W_{10}(y)}dy}\frac{E_1}
{\chi_{1\to 3}(\lambda)\chi_{3\to {\bar 3}}(\lambda)}
\frac{\chi_3(\zeta_2^{(1)}(\lambda),\lambda)}{\chi_4(\zeta_2^{(1)}(\lambda),\lambda)}
\label{36}
\ee
from which we get infinitely many solutions for $\zeta_2^{(1)}(\lambda)$:
\be
\int_{z_2}^{\zeta_{2,m}^{(1)}(\lambda)}\sqrt{W_{10}(y)}dy=-(m-\frac{1}{4})\frac{i\pi}{\lambda}+
\fr\oint_{K_1}\sqrt{W_{10}(y)}dy+
\frac{1}{2\lambda}\ln E_1+\nn\\
\frac{1}{2\lambda}\ln\ll(\frac{1}{\chi_{1\to 3}(\lambda)\chi_{3\to {\bar 3}}(\lambda)}
\frac{\chi_3(\zeta_{2,m}^{(1)}(\lambda),\lambda)}{\chi_4(\zeta_{2,m}^{(1)}(\lambda),\lambda)}\r)
\label{361}
\ee
with integer $m$.

Taking therefore the regular limit $\lambda_{s_1}\to\infty$ ($R_1$ is fixed) we get:
\be
\int_{K_2(\zeta_{2,qr}^{(1)}(\lambda_{s_1}))}\ll(\fr\sqrt{W_{10}(y)}+
\frac{1}{2\lambda_{s_1}}Z^-(y,\lambda_{s_1})\r)dy=(q[\lambda_{s_1}]+r-\frac{1}{4}-\frac{R_1}{2})\frac{i\pi}{\lambda_{s_1}}-\nn\\
\frac{1}{4\lambda_{s_1}}\oint_{K_1}Z^-(y,\lambda_{s_1})dy-
\frac{1}{2\lambda_{s_1}}\ln(2\cos(R_1\pi-\frac{i}{2}\oint_{K_1}Z^-(y,\lambda_{s_1})dy))\nn\\
q=0,1,2,...,\;r=0,1,...,[\lambda_{s_1}]-1
\label{363}
\ee

It is seen therefore that similarly to the critical case of SG considered earlier (see
{\bf Theorem 1b}) zeros of $\psi_1(z,\lambda)$ are distributed along the infinite SL emerging from the turning
point $z_2$.

{\it The case 4.}

In fact, by considering this case we would like to check that the results of the previous case are valid in a vicinity
of the inner SL between $z_6$ and $z_{\bar 6}$, where the combination in the second of eq. \mref{34} can not be continued
canonically. We can use the first of the eq. \mref{34} to get:
\be
\psi_1(z,\lambda)=\ll(\alpha_{\frac{1}{3}\to{\bar 3}}+\alpha_{\frac{1}{\bar 3}\to{3}}\alpha_{\frac{\bar 3}{3}\to{5}}\r)\psi_3(z,\lambda)+
\alpha_{\frac{1}{\bar 3}\to{3}}\alpha_{\frac{\bar 3}{5}\to{3}}\psi_5(z,\lambda)=\nn\\
-iE_2\frac{\chi_3(z,\lambda)}{\chi_{3\to {\bar 3}}(\lambda)}e^{\ll(-\lambda\int_{z_2}^{z}+\fr\lambda\oint_{K_1}\r)\sqrt{W_{10}(y)}dy}+
\frac{\chi_{1\to 3}(\lambda)\chi_5(z,\lambda)}{\chi_{3\to 5}(\lambda)}e^{\ll(\lambda\int_{z_2}^{z}-\fr\lambda\oint_{K_1}\r)\sqrt{W_{10}(y)}dy}
\label{364}
\ee
where
\be
E_2=\chi_{1\to{\bar 3}}(\lambda)+
e^{-\lambda\oint_{K_1}\sqrt{W_{10}(y)}dy}\chi_{1\to{3}}(\lambda)\frac{\chi_{{\bar 3}\to 5}(\lambda)}{\chi_{3\to 5}(\lambda)}
\label{365}
\ee

Therefore from \mref{364} we get:
\be
\int_{z_2}^{\zeta_{2,m}^{(1)}(\lambda)}\sqrt{W_{10}(y)}dy=-(m-\frac{1}{4})\frac{i\pi}{\lambda}+
\fr\oint_{K_1}\sqrt{W_{10}(y)}dy+
\frac{1}{2\lambda}\ln E_2+\nn\\
\frac{1}{2\lambda}\ln\ll(\frac{\chi_{3\to 5}(\lambda)}{\chi_{1\to 3}(\lambda)\chi_{3\to {\bar 3}}(\lambda)}
\frac{\chi_3(\zeta_{2,m}^{(1)})(\lambda),\lambda)}{\chi_5(\zeta_{2,m}^{(1)}(\lambda),\lambda)}\r)
\label{366}
\ee
with integer $m$.

Taking now the limit $\lambda_{s_1}\to\infty$ in the last formula we obtain again the result \mref{363}.

{\it The case 5.}

For this case we have:
\be
\psi_1(z,\lambda)=(\alpha_{\frac{1}{3}\to{\bar 3}}\alpha_{\frac{3}{4}\to 5}+
\alpha_{\frac{1}{\bar 3}\to{3}}\alpha_{\frac{\bar 3}{4}\to 5})\psi_4(z,\lambda)+\nn\\
(\alpha_{\frac{1}{3}\to{\bar 3}}\alpha_{\frac{3}{5}\to 4}+
\alpha_{\frac{1}{\bar 3}\to{3}}\alpha_{\frac{\bar 3}{5}\to 4})\psi_5(z,\lambda)=\nn\\
\frac{W_{10}^{-\frac{1}{4}}(z)
e^{\ll(-\lambda\int_{z_2}^{z_6}+\fr\lambda\oint_{K_1}\r)\sqrt{W_{10}(y)}dy}}{\chi_{3\to{\bar 3}}(\lambda)}\times\nn\\
\ll(-i\chi_{3\to 5}(\lambda)E_2\chi_4(z,\lambda)e^{-\lambda\int_{z_6}^{z}\sqrt{W_{10}(y)}dy}+
E_1\chi_5(z,\lambda)e^{\lambda\int_{z_6}^{z}\sqrt{W_{10}(y)}dy}\r)
\label{37}
\ee

Let us notice that:
\be
E_2=E_1+
e^{2\lambda\int_{z_2}^{z_6}\sqrt{W_{10}(y)}dy-\lambda\oint_{K_1}\sqrt{W_{10}(y)}dy}
\chi_{1\to{3}}(\lambda)\frac{\chi_{3\to{\bar 3}}(\lambda)}{\chi_{3\to 5}(\lambda)}
\label{371}
\ee

To get the last equation we have made use of the following identity:
\be
\frac{\chi_{{\bar 3}\to 5}(\lambda)}{\chi_{3\to 5}(\lambda)}=\chi_{\bar 3\to 4}(\lambda)+
e^{2\lambda\int_{z_2}^{z_6}\sqrt{W_{10}(y)}dy}
\frac{\chi_{3\to{\bar 3}}(\lambda)}{\chi_{3\to 5}(\lambda)}
\label{39}
\ee

The above identity is typical for each four FS's communicating with themselves canonically (see ref.1,2 of \cite{5} and
App. B, formula \mref{B6}). Here these FS's are $\psi_i(z,\lambda)$ with $i=3,{\bar 3},4,5$.

Now from \mref{37} for the distribution of zeros $\zeta_{6,m}^{(1)}$ of $\psi_1(z)$
in a vicinity of the right SL emerging from $z_6$ we get:
\be
\int_{z_6}^{\zeta_{6,m}^{(1)}(\lambda)}\sqrt{W_{10}(y)}dy=-(m-\frac{1}{4})\frac{i\pi}{\lambda}+\nn\\
\frac{1}{2\lambda}\ln\frac{E_2}{E_1}+\ln\frac{\chi_{3\to 5}(\lambda)\chi_4(\zeta_{6,m}^{(1)}(\lambda),\lambda)}{\chi_5(\zeta_{6,m}^{(1)}(\lambda),\lambda)}
\label{38}
\ee

Taking now into account the relation \mref{371} we get from the above formula for the regular limit
$\lambda(=[\lambda]+\Lambda)\to+\infty,\;\Lambda$ - fixed, the following result:
\be
\int_{K_6(\zeta_{6,qr}^{(1)}(\lambda))}\ll(\fr\sqrt{W_{10}(y)}+\frac{1}{2\lambda}Z^-(y,\lambda)\r)dy=
-(q[\lambda]+r-\frac{1}{4})\frac{i\pi}{\lambda}\nn\\
q=0,1,2,3,...,\;r=0,1,...,[\lambda]-1
\label{40}
\ee
i.e. the limit loci of zeros $\zeta_{6,m}^{(1)}(\lambda)$ of $\psi_1(z,\lambda)$ in the case considered is just the right SL emerging from
the turning point $z_6$ (see Fig.5).

{\it The case 6.}

For this case we have:
\be
\psi_1(z,\lambda)=A\psi_5(z,\lambda)+{\bar A}\psi_{\bar 5}(z,\lambda)=\nn\\
W_{10}^{-\frac{1}{4}}(y)\ll(A\chi_5(z,\lambda)e^{\lambda\int_{z_4}^{z}\sqrt{W_{10}(y)}dy}+
          i{\bar A}\chi_{\bar 5}(z,\lambda)e^{-\lambda\int_{z_4}^{z}\sqrt{W_{10}(y)}dy}\r)
\label{41}
\ee
where $A=\alpha_{\frac{1}{3}\to{\bar 3}}\alpha_{\frac{3}{5}\to{\bar 5}}+
\alpha_{\frac{1}{\bar 3}\to{3}}\alpha_{\frac{\bar 3}{5}\to{\bar 5}}$ and the last line in \mref{41} has been written
above the cut between $z_3$ and $z_4$.

By its definition $A$ is given by:
\be
A=-E_3
\frac{\chi_{3\to{\bar 5}}(\lambda)}{\chi_{3\to{\bar 3}}(\lambda)\chi_{5\to{\bar 5}}(\lambda)}
e^{-\lambda\ll(\int_{z_2}^{z_3}+\fr\oint_{K_2}-\fr\oint_{K_1}\r)\sqrt{W_{10}(y)}dy}
\label{42}
\ee
where
\be
E_3=\chi_{1\to{\bar 3}}+
e^{-\lambda\oint_{K_1}\sqrt{W_{10}(y)}dy}\chi_{1\to 3}(\lambda)\frac{\chi_{{\bar 3}\to{\bar 5}}(\lambda)}{\chi_{3\to{\bar 5}}(\lambda)}=E_1+\nn\\
e^{2\lambda\int_{z_2}^{z_6}\sqrt{W_{10}(y)}dy-\lambda\oint_{K_1}\sqrt{W_{10}(y)}dy}
\chi_{1\to 3}(\lambda)\frac{\chi_{3\to{\bar 3}}(\lambda)\chi_{4\to{\bar 5}}(\lambda)}{\chi_{3\to{\bar 5}}(\lambda)}
\label{421}
\ee

The last equation in \mref{421} has been obtained by using the identity:
\be
\frac{\chi_{{\bar 3}\to{\bar 5}}(\lambda)}{\chi_{3\to{\bar 5}}(\lambda)}=\chi_{\bar 3\to 4}(\lambda)+
e^{2\lambda\int_{z_2}^{z_6}\sqrt{W_{10}(y)}dy}\frac{\chi_{3\to{\bar 3}}(\lambda)\chi_{4\to{\bar 5}}(\lambda)}{\chi_{3\to{\bar 5}}(\lambda)}
\label{44}
\ee

A condition for the distribution of zeros $\zeta_{3,m}^{(1)}$ of $\psi_1(z)$ in the vicinity of the internal SL linking
$z_3$ with $z_4$ takes therefore the form (above the cut $z_3,z_4$):
\be
\int_{z_3}^{\zeta_{3,m}^{(1)}}\sqrt{W_{10}(y)}dy=(m-\frac{1}{4})\frac{i\pi}{\lambda}-
\fr\oint_{K_1}\sqrt{W_{10}(y)}dy+\frac{1}{2\lambda}\ln\frac{{\bar E}_3}{E_3}+\nn\\
\frac{1}{2\lambda}\ln\frac{\chi_{{\bar 3}\to 5}(\lambda)\chi_{\bar 5}(\zeta_{3,m}^{(1)},\lambda)}{\chi_{3\to{\bar 5}}(\lambda)\chi_5(\zeta_{3,m}^{(1)},\lambda)}
\label{43}
\ee

Taking into account the definition of $E_3$ by \mref{421} we see that in the regular limit $\lambda(=[\lambda]+\Lambda)
\to\infty,\;\Lambda$ - fixed, the distribution of zeros of $\psi_1(z,\lambda)$ is given in this case by the condition:
\be
\int_{K_3{(\zeta_{3,qr}^{(1)}}(\lambda))}\ll(\fr\sqrt{W_{10}(y)}dy+\frac{1}{2\lambda}
Z^-(y,\lambda)dy\r)=-(q[\lambda]+r-\frac{1}{4})\frac{i\pi}{\lambda}\nn\\
q=0,1,2,...,q_2,\;\;r=0,1,...,[\lambda]-1
\label{45}
\ee
where $q_2\geq 0$ is given by $\int_{z_3}^{z_4}\sqrt{W_{10}(y)}dy=(q_2+r_2)i\pi,\;0\leq r_2<1$.

The loci of zeros $\zeta_{3,m}^{(1)}(\lambda)$ of $\psi_1(z,\lambda)$ given by \mref{45} is therefore the internal SL
between $z_3$ and $z_4$.

{\it The case 7.}

The linear combination is now the following:
\be
\psi_1(z,\lambda)=(A+{\bar A}\alpha_{\frac{\bar 5}{5}\to 6})\psi_5(z,\lambda)+
{\bar A}\alpha_{\frac{\bar 5}{6}\to 5}\psi_6(z,\lambda)=\nn\\
W_{10}^{-\frac{1}{4}}(y)\ll[(A+{\bar A}\chi_{{\bar 5}\to 6}(\lambda))\chi_5(z,\lambda)
e^{\lambda\int_{z_4}^{z}\sqrt{W_{10}(y)}dy}+\r.\nn\\
         \ll. i{\bar A}\chi_{5\to{\bar 5}}(\lambda)\chi_6(z,\lambda)e^{-\lambda\int_{z_4}^{z}\sqrt{W_{10}(y)}dy}\r]
\label{46}
\ee
so that for the corresponding distribution of zeros $\zeta_{4,m}^{(1)}(\lambda)$ we get:
\be
\int_{z_4}^{\zeta_{4,m}^{(1)}}\sqrt{W_{10}(y)}dy=(m-\frac{1}{4})\frac{i\pi}{\lambda}-
\frac{1}{2\lambda}\ln\ll(\frac{A}{\bar A}+\chi_{{\bar 5}\to 6}\r)+
\frac{1}{2\lambda}\ln\frac{\chi_{{\bar 5}\to 5}\chi_6(\zeta_{4,m}^{(1)})}{\chi_5(\zeta_{4,m}^{(1)})}
\label{47}
\ee

Using again both the formulae \mref{42} and \mref{421} we get from \mref{47} for the regular limit
$\lambda_{s_2}\to\infty$:
\be
\int_{K_4(\zeta_{4,qr}^{(1)}(\lambda_{s_2}))}\ll(\fr\sqrt{W_{10}(y)}+
\frac{1}{2\lambda_{s_2}}Z^-(y,\lambda_{s_2})\r)dy=-(q[\lambda_{s_2}]+r-\frac{1}{4}-
\frac{R_2}{2})\frac{i\pi}{\lambda_{s_2}}-\nn\\
\frac{1}{4\lambda_{s_2}}\oint_{K_2}Z^-(y,\lambda_{s_2})dy+
\frac{1}{2\lambda_{s_2}}\ln(2\cos(R_2\pi-\frac{i}{2}\oint_{K_2}Z^-(y,\lambda_{s_2})dy))\nn\\
q=0,1,2,...,\;r=0,1,...,[\lambda_{s_2}]-1
\label{48}
\ee

It follows from \mref{48} that the zeros $\zeta_{4,m}^{(1)}(\lambda_{s_2})$ all lie along the infinite SL emerging from
$z_4$.

{\it The case 8.}

Expressing $\psi_5(z,\lambda)$ in the formula \mref{46} by $\psi_6(z,\lambda)$ and $\psi_7(z,\lambda)$ we get:
\be
\psi_1(z,\lambda)=[(A+{\bar A}\alpha_{\frac{\bar 5}{5}\to 6})\alpha_{\frac{5}{6}\to 7}+
{\bar A}\alpha_{\frac{\bar 5}{6}\to 5}]\psi_6(z,\lambda)+
(A+{\bar A}\alpha_{\frac{\bar 5}{5}\to 6})\alpha_{\frac{5}{7}\to 6}\psi_7(z,\lambda)
\label{49}
\ee
and the condition for zeros $\zeta_{7,m}^{(1)}(\lambda)$ of $\psi_1(z,\lambda)$ takes on the form:
\be
\int_{z_7}^{\zeta_{7,m}^{(1)}(\lambda)}\sqrt{W_{10}(y)}dy=(m-\frac{1}{4})\frac{i\pi}{\lambda}-
\frac{1}{2\lambda}\ln\ll(\chi_{5\to 7}(\lambda)+
\frac{\chi_{5\to{\bar 5}}(\lambda)e^{-2\lambda\int_{z_4}^{z_7}\sqrt{W_{10}(y)}dy}}{\frac{A}{\bar A}+\chi_{{\bar 5}\to 6}(\lambda)}\r)+\nn\\
\frac{1}{2\lambda}\ln\frac{\chi_7(\zeta_{7,m}^{(1)}(\lambda),\lambda)}{\chi_6(\zeta_{7,m}^{(1)}(\lambda),\lambda)}
\label{50}
\ee

The asymptotic regular limit $\lambda(=\lambda]+\Lambda)\to\infty$ with fixed $\Lambda$ which we get from \mref{50} is
therefore the following:
\be
\int_{K_7(\zeta_{7,qr}^{(1)}(\lambda))}\ll(\fr\sqrt{W_{10}(y)}+
\frac{1}{2\lambda}Z^-(y,\lambda)\r)dy=(q[\lambda]+r-\frac{1}{4})\frac{i\pi}{\lambda}\nn\\
q=0,1,2,...,\;r=0,1,...,[\lambda]-1
\label{51}
\ee
i.e. all zeros $\zeta_{7,m}^{(1)}(\lambda)$ of $\psi_1(z,\lambda)$ tend to lie on the infinite SL emerging from $z_7$ and being a part
of the sector $S_7$ boundary.

Since the rest of the asymptotic distribution of zeros of $\psi_1(z,\lambda)$ can be obtained by the complex conjugation of loci
of zeros just established in the distinguished cases 1.-8. above the final picture of their loci on the
$C_{cut}$-plane is shown in Fig.4a as the bold lines.

\vskip 15pt

\begin{tabular}{c}
\psfig{figure=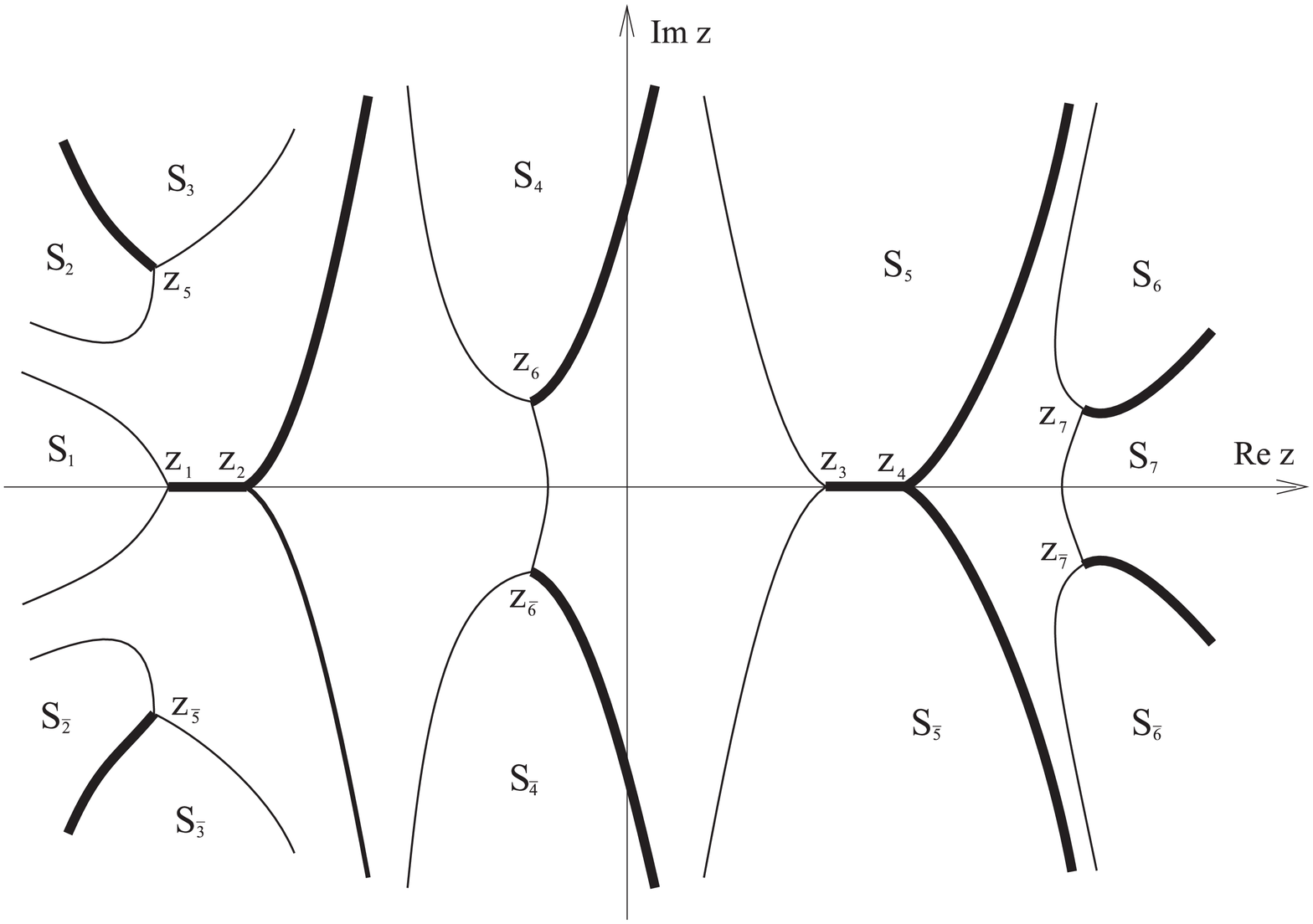,width=11cm}\\
Fig.4a ESL's (bold Stokes lines) for $\psi_1(z,\lambda)$ in the regular limits $\lambda\to\infty$.\\
The non-quantized case
\end{tabular}

\vskip 15pt

We can perform analogous calculation looking for zeros of $\psi_7(z,\lambda)$ in the limit $\lambda\to\infty$ and {\it mutatis
mutandis} we get the results drawn in Fig.4b as the bold SL's.

Let us note also in a relation with the next sections that the obtained two patterns of the limit loci of zeros of the FS's
$\psi_1(z,\lambda)$ and $\psi_7(z,\lambda)$ do not change essentially if the double well considered is symmetric. These patterns get
simply additional property to be mutually symmetric i.e. they can be obtained from each other by the inversion operation
$z\to-z$.

\section{The quantized asymmetric double--well potential}

\hskip+2em Let us consider now possible changes which quantization of $\lambda$ can cause in the above distribution
pictures of zeros of $\psi_1(z,\lambda)$. We quantize $\lambda$ by matching $\psi_1(z,\lambda)$ with $\psi_7(z,\lambda)$.
By this condition both the solution are in fact identified and one can expect that the two separate
distributions shown in Fig.Fig.4a-4b are also unified somehow. It is clear that to get figures corresponding to such a
unified configuration of zeros one has to remove some ESL's from the figures as well as to add some of them. However one
needs to know rules governing such a procedure and the goal of the detailed calculations below is to provide us with the
rules.

We shall start matching the solutions $\psi_1(z,\lambda)$ and $\psi_7(z,\lambda)$. We can do it using
the combination \mref{49} and putting equal to zero the coefficient at $\psi_6(z,\lambda)$. We get:
\be
(A+{\bar A}\alpha_{\frac{\bar 5}{5}\to 6})\alpha_{\frac{5}{6}\to 7}+
{\bar A}\alpha_{\frac{\bar 5}{6}\to 5}=\alpha_{\frac{5}{6}\to 7}(A+{\bar A}\alpha_{\frac{\bar 5}{5}\to 7})=\nn\\
\frac{\alpha_{\frac{5}{6}\to 7}}{\chi_{5\to 7}}(A\chi_{5\to 7}+{\bar A}\chi_{{\bar 5}\to 7})=0
\label{52}
\ee

\vskip 15pt

\begin{tabular}{c}
\psfig{figure=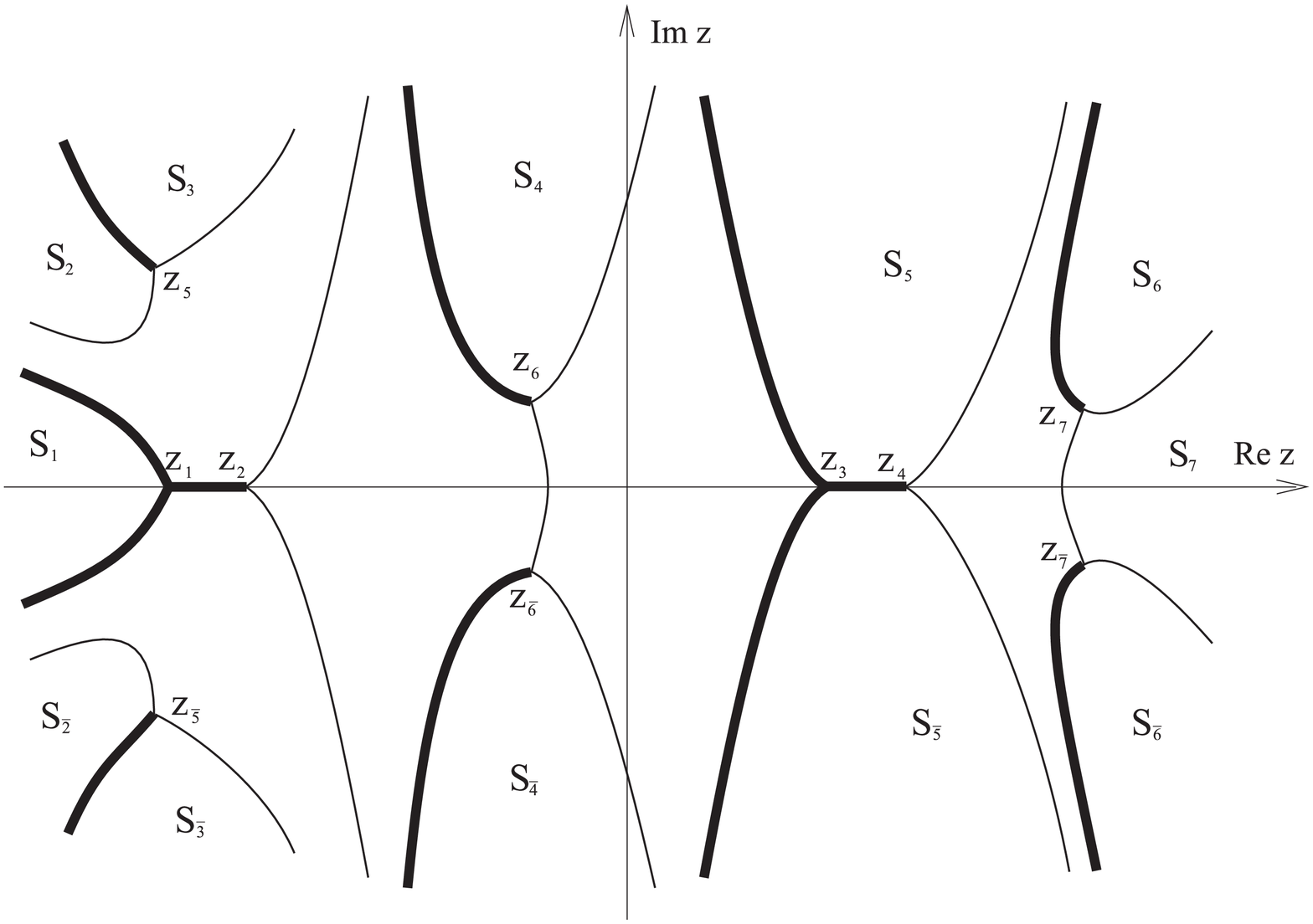,width=11cm}\\
Fig.4b ESL's (bold Stokes lines) of $\psi_7(z,\lambda)$ in the regular limits $\lambda\to\infty$.\\
The non-quantized case
\end{tabular}

\vskip 15pt

Taking into account \mref{42} we obtain the following form of the condition for the $\lambda$-quantization:
\be
E_3(\lambda)\chi_{5\to 7}(\lambda)\chi_{3\to{\bar 5}}(\lambda)e^{-\lambda\oint_{K_2}\sqrt{W_{10}(y)}dy}+
E_2(\lambda)\chi_{{\bar 5}\to 7}(\lambda)\chi_{3\to 5}(\lambda)=0
\label{31}
\ee

The above quanization condition can be further elaborated with the help of the relation \mref{371} and \mref{421} to get:
\be
E_1E_4=E_1\frac{\chi_{5\to 7}\chi_{5\to{\bar 5}}}{\chi_{3\to 5}}e^{2\lambda\int_{z_6}^{z_3}\sqrt{W_{10}(y)}dy-
\lambda\oint_{K_2}\sqrt{W_{10}(y)}dy}-\nn\\
E_4\frac{\chi_{1\to 3}\chi_{3\to{\bar 3}}}{\chi_{3\to 5}}e^{2\lambda\int_{z_2}^{z_6}\sqrt{W_{10}(y)}dy-
\lambda\oint_{K_1}\sqrt{W_{10}(y)}dy}
\label{33}
\ee
where
\be
E_4(\lambda)=\chi_{{\bar 5}\to 7}(\lambda)+
e^{-\lambda\oint_{K_2}\sqrt{W_{10}(y)}dy}\chi_{5\to 7}(\lambda)\chi_{4\to{\bar 5}}(\lambda)
\label{53}
\ee
The formula \mref{33} is the {\it exact} quantization formula valid for {\it any} real D-W polynomial potential
with the properly chosen FS's $\psi_i(z),\;i=3,{\bar 3},5,{\bar 5}$.

The following equivalent form of the quantization condition \mref{52} will appear also to be useful in our further
analysis:
\be
\frac{A}{\bar A}+{\chi_{\bar 5\to 6}}=-\frac{\chi_{5\to{\bar 5}}}{\chi_{5\to 7}}
e^{-2\lambda\int_{z_4}^{z_7}\sqrt{W_{10}(y)}dy}
\label{54}
\ee
which follows directly from \mref{52} when the indentity:
\be
\frac{\chi_{{\bar 5}\to 7}}{\chi_{5\to 7}}={\chi_{\bar 5\to 6}}+
\frac{\chi_{5\to{\bar 5}}}{\chi_{5\to 7}}e^{-2\lambda\int_{z_4}^{z_7}\sqrt{W_{10}(y)}dy}
\label{55}
\ee
is used.

Consider now the $\lambda\to\infty$-limit of the above quantization formulae up to {\it any} order
of $\lambda^{-1}$.

Using the representations \mref{551}-\mref{554} for the semiclassical expansion of the $\chi$-factors we get for
the semiclassical limit of the quantization formula \mref{33}, the following result:
\be
E_1^{as}(\lambda)E_4^{as}(\lambda)=\ll[1+\exp\ll(-\lambda\oint_{K_1}\sqrt{W_{10}(y)}dy-
\sum_{n{\geq}0}\left(\frac{1}{2\lambda}\right)^{2n+1}\oint_{K_1}X_{2n+1}(y)dy\r)\r]\times
\nn\\
\ll[1+\exp\ll(-\lambda\oint_{K_2}\sqrt{W_{10}(y)}dy-
\sum_{n{\geq}0}\left(\frac{1}{2\lambda}\right)^{2n+1}\oint_{K_2}X_{2n+1}(y)dy\r)\r]=0
\label{56}
\ee
what means that asymptotically $\lambda(>0)$ is quantized in each well independently being defined in the respective wells by the following
conditions:
\be
\lambda_r^{(l)}\oint_{K_l}\sqrt{W_{10}(y)}dy+
\sum_{n{\geq}0}\left(\frac{1}{2\lambda_r^{(l)}}\right)^{2n+1}\oint_{K_l}X_{2n+1}(y)dy=(-1)^{l+1}(2r+1)i\pi\nn\\
\lambda_r^{(l)}\in\Lambda_l,\;l=1,2
\label{57}
\ee
while $r$ is natural and large, i.e. $r\gg 1$.

Comparing \mref{57} with \mref{261}-\mref{262} it is seen that $R=\fr$ for both $l=1,2$, i.e. the limits
$\lambda_r^{(l)}\to\infty$ are singular what needs a special care in considering these limits.

The explicite solutions to the above quantization conditions can be obtained by iterations to
get the following forms:
\be
\lambda_r^{(l)}=(-1)^{l+1}(2r+1)\frac{i\pi}{\oint_{K_l}\sqrt{W_{10}(y)}dy}-
\sum_{n{\geq}0}\left(\frac{1}{2\lambda_r^{(l)}}\right)^{2n+1}\frac{\oint_{K_l}X_{2n+1}(y)dy}
{\oint_{K_l}\sqrt{W_{10}(y)}dy}=\nn\\
(-1)^{l+1}(2r+1)\frac{i\pi}{\oint_{K_l}\sqrt{W_{10}(y)}dy}+\sum_{n{\geq}1}\frac{a_n^{(l)}}{(2r+1)^n}\nn\\
\;\;\;\;l=1,2,\;\;r=0,1,2,...
\label{571}
\ee
with the following first three coefficients:
\be
a_1^{(l)}=\frac{(-1)^l}{2\pi i}\oint_{K_l}X_1(y)dy\nn\\
a_2^{(l)}=0\nn\\
a_3^{(l)}=-\frac{(-1)^l}{(i\pi)^3}\oint_{K_l}\sqrt{W_{10}(y)}dy\ll(\ll(\oint_{K_l}X_1(y)dy\r)^2+\r.\nn\\
\ll.\oint_{K_l}\sqrt{W_{10}(y)}dy\oint_{K_l}X_3(y)dy\r),\;\;\;l=1,2
\label{572}
\ee

While a usefullness of the formulae \mref{571} as the asymptotic ones is for large $r$ the formulae themselves can be
considered for any $r\geq 0$ and this last range for $r$ will be assumed in our further considerations.

It is obvious from the form of the expansions \mref{571} that the spectra $\Lambda_l,\;l=1,2$
can coincide only for the symmetric D-W polynomial potentials since in other cases there are
no a sufficient number of coefficients of the polynomial $W_{10}(z)$ to satisfy all the equations
$a_n^{(1)}=a_n^{(2)},\;n\geq 1$. It means that these spectra can coincide on some of their part
only up to some order. This can happen for example when the following conditions are satisfied:
\be
\frac{\oint_{K_1}\sqrt{W_{10}(y)}dy}{\oint_{K_2}\sqrt{W_{10}(y)}dy}=
\frac{\oint_{K_1}X_1(y)dy}{\oint_{K_2}X_1(y)dy}=...=
\frac{\oint_{K_1}X_{2k-1}(y)dy}{\oint_{K_2}X_{2k-1}(y)dy}=-\frac{2p+1}{2q+1}\nn\\
p,q=0,1,...,\;p\neq q,\;\;\;0\leq k<10
\label{573}
\ee
i.e. the equalities $a_n^{(1)}=a_n^{(2)},\;n\geq 1,$ are then satisfied up to $n=2k-1$.

For these cases the spectra have subsequences consisiting of
$\lambda_{(2p+1)r+p}^{(1)}$ and $\lambda_{(2q+1)r+q}^{(2)}$, $r=0,1,2...$, for which their expansions \mref{571} coincide
up to $2k+1$-th order. Using \mref{573} we have then:
\be
\lambda_{(2q+1)r+q}^{(2)}\oint_{K_1}\sqrt{W_{10}(y)}dy+
\sum_{n{\geq}0}\left(\frac{1}{2\lambda_{(2q+1)r+q}^{(2)}}\right)^n\oint_{K_1}X_{2n+1}(y)dy=\nn\\
-(2(2p+1)r+2p+1)i\pi+
\sum_{n>k}\left(\frac{1}{2\lambda_{(2q+1)r+q}^{(2)}}\right)^n\ll(\oint_{K_1}+
\frac{2p+1}{2q+1}\oint_{K_2}\r)X_{2n+1}(y)dy
\label{5731}
\ee
and {\it mutatis mutandis}
\be
\lambda_{(2p+1)r+p}^{(1)}\oint_{K_2}\sqrt{W_{10}(y)}dy+
\sum_{n{\geq}0}\left(\frac{1}{2\lambda_{(2p+1)r+p}^{(1)}}\right)^n\oint_{K_2}X_{2n+1}(y)dy=\nn\\(2(2q+1)r+2q+1)i\pi+
\sum_{n>k}\left(\frac{1}{2\lambda_{(2p+1)r+p}^{(1)}}\right)^n\ll(\oint_{K_2}+
\frac{2q+1}{2p+1}\oint_{K_1}\r)X_{2n+1}(y)dy
\label{5732}
\ee

We can conclude therefore that for asymmetric D-W polynomial potentials considering the limit $\lambda\to\infty$ we have
to take into account only sequences $\{\lambda_r^{(l)},\;r=1,2,...\},\;l=1,2$, consisting of not coinciding spectra
or the subsequences just considered which we denote by
$\{\lambda_s^{(3)},\;s=1,2,...\}$, so that $\{\lambda_s^{(3)}\}\equiv\{\lambda_{(2p+1)s+p}^{(1)}\}$ or
$\{\lambda_s^{(3)}\}\equiv\{\lambda_{(2q+1)s+q}^{(2)}\}$.

Let us now discuss changes which the above different cases of the $\lambda$-quantization conditions can
introduce to the $\lambda\to\infty$-limit zero distributions of $\psi_1(z,\lambda)$ (as well as of $\psi_7(z,\lambda)$
this time) considered
in the previous section. In our analysis we have to be particularly careful about
these formulae of the previous section which contain as the $\log$-function arguments terms
which limits are the factors of the limit quantization formula \mref{56} or the factors leading to the formulae
\mref{5731}-\mref{5732}.

We shall consider all the cases of the previous section subsequently.

{\it The case} $1.'$

The $\lambda$-quantization for this case does not disturb the condition \mref{272} for zeros distribution of
$\psi_1(z,\lambda)$ so that its $\lambda_s^{(i)}\to\infty$-limit, i.e. for $s\to\infty$, also remains unchanged
independently of $i=1,2,3$ and we get readily:
\be
\int_{K_5{(\zeta_{5,qr}^{(1)}}(\lambda_s^{(i)}))}\ll(\fr\sqrt{W_{10}(y)}dy+\frac{1}{2\lambda_s^{(i)}}
Z^-(y,\lambda_s^{(i)})dy\r)=(q[\lambda_s^{(i)}]+r-\frac{1}{4})\frac{i\pi}{\lambda_s^{(i)}}\nn\\
q=0,1,2,3,...,\;r=0,1,...,[\lambda_s^{(i)}]-1
\label{574}
\ee

{\it The case} $2.'$

Also in this case the $\lambda$-quantization does change almost nothing in the $\lambda\to\infty$-limit distribution of zeros
except that it fixes $R_1$ on $\fr$ for the sequence $\lambda_s^{(1)}\to\infty$ while $R_1$ depends on $s$ for the
sequence $\lambda_s^{(2)}\to\infty$, i.e we have:
\be
\int_{K_1{(\zeta_{1,qr}^{(1)}}(\lambda_s^{(i)}))}\ll(\fr\sqrt{W_{10}(y)}dy+\frac{1}{2\lambda_s^{(i)}}
Z^-(y,\lambda_s^{(i)})dy\r)=(q[\lambda_s^{(i)}]+r-\frac{1}{4})\frac{i\pi}{\lambda_s^{(i)}}\nn\\
q=0,1,2,...,q_1,\;r=0,1,...,[\lambda_s^{(i)}]-1
\label{575}
\ee
with integer $q_1\geq 0$ given by $\int_{z_1}^{z_2}\sqrt{W_{10}(y)}dy=-(q_1+r_1)i\pi,\;0\leq r_1<1$.

{\it The case} $3.'$

In this case the quantization of $\lambda$ can disturb only the term $\ln E_1$ present in the r.h.s. of the \mref{361}
when $\lambda$ is quantized in the first well.

To estimate its $\lambda\to\infty$-behaviour let us note that the quantization condition can be also written in the form:
\be
E_1E_5=-e^{2\lambda\int_{z_2}^{z_6}\sqrt{W_{10}(y)}dy-\lambda\oint_{K_1}\sqrt{W_{10}(y)}dy}
       \frac{\chi_{1\to 3}\chi_{3\to{\bar 3}}}{\chi_{3\to 5}}E_4\nn\\
\label{58}
\ee
where
\be
E_5=\chi_{{\bar 5}\to 7}+e^{-\lambda\oint_{K_2}\sqrt{W_{10}(y)}dy}\chi_{5\to{7}}\frac{\chi_{3\to{\bar 5}}}{\chi_{3\to 5}}
=\nn\\
E_4-\frac{\chi_{5\to 7}\chi_{5\to{\bar 5}}}{\chi_{3\to 5}}e^{2\lambda\int_{z_6}^{z_3}\sqrt{W_{10}(y)}dy-
\lambda\oint_{K_2}\sqrt{W_{10}(y)}dy}
\label{59}
\ee
where the second equation has been obtained due to the following identity:
\be
\frac{\chi_{3\to{\bar 5}}}{\chi_{3\to 5}}=\chi_{{\bar 5\to 4}}-
e^{2\lambda\int_{z_6}^{z_3}\sqrt{W_{10}(y)}dy}\frac{\chi_{5\to{\bar 5}}}{\chi_{3\to 5}}
\label{60}
\ee

From \mref{58} and \mref{59} we get:
\be
\frac{1}{E_1}=-e^{\ll(-2\lambda\int_{z_2}^{z_6}+\lambda\oint_{K_1}\r)\sqrt{W_{10}(y)}dy}
\frac{\chi_{3\to 5}}{\chi_{1\to 3}\chi_{3\to{\bar 3}}}\frac{E_5}{E_4}=\nn\\
-e^{\ll(-2\lambda\int_{z_2}^{z_6}+\lambda\oint_{K_1}\r)\sqrt{W_{10}(y)}dy}
\frac{\chi_{3\to 5}}{\chi_{1\to 3}\chi_{3\to{\bar 3}}}\times\nn\\
\ll(1-\frac{\chi_{5\to 7}\chi_{5\to{\bar 5}}}{\chi_{3\to 5}E_4}e^{2\lambda\int_{z_6}^{z_3}\sqrt{W_{10}(y)}dy-
\lambda\oint_{K_2}\sqrt{W_{10}(y)}dy}\r)
\label{61}
\ee

Now we take in \mref{61} the limit $\lambda_s^{(1)}\to\infty$. Therefore $E_4(\lambda_s^{(1)})\neq 0$ in \mref{61} and
we get in this limit for $\ln E_1$:
\be
\ln E_1^{as}=-\ln\frac{1}{E_1^{as}}=\pm i\pi+2\lambda_s^{(1)}\int_{z_2}^{z_6}\sqrt{W_{10}(y)}dy-
\lambda_s^{(1)}\oint_{K_1}\sqrt{W_{10}(y)}dy-\nn\\
\ln\ll(\frac{\chi_{3\to 5}}{\chi_{1\to 3}\chi_{3\to{\bar 3}}}\r)^{as}
\label{62}
\ee

Substituting the last result to \mref{361} we get for $\lambda$'s quantized in the left well:
\be
\int_{K_6(\zeta_{6,qr}^{(1)}(\lambda_s^{(1)}))}\ll(\fr\sqrt{W_{10}(y)}+
\frac{1}{2\lambda_s^{(1)}}Z^-(y,\lambda_s^{(1)})\r)dy=(q[\lambda_s^{(1)}]+r-\frac{1}{4})\frac{i\pi}{\lambda_s^{(1)}}\nn\\
q=0,1,2,...,\;r=0,1,...,[\lambda_s^{(1)}]-1
\label{63}
\ee

The result \mref{63} is essentially different from \mref{363} since now zeros $\zeta_{2,m}^{(1)}(\lambda_{s_1})$
of $\psi_1(z,\lambda)$ have all
been shifted to the positions $\zeta_{6,m}^{(1)}(\lambda_s^{(1)})$ on the left infinite SL emerging from the turning point
$z_6$.

Let us note however that the distribution defined by \mref{363} is kept unchanged if
the limit $\lambda_s^{(2)}\to\infty$ is taken so that we have for this case:
\be
\int_{K_2(\zeta_{2,qr}^{(1)}(\lambda_s^{(2)}))}\ll(\fr\sqrt{W_{10}(y)}+
\frac{1}{2\lambda_s^{(2)}}Z^-(y,\lambda_s^{(2)})\r)dy=(q[\lambda_s^{(2)}]+r-\frac{1}{4}-
\frac{R_1}{2})\frac{i\pi}{\lambda_s^{(2)}}-\nn\\
\frac{1}{4\lambda_s^{(2)}}\oint_{K_1}Z^-(y,\lambda_s^{(2)})dy-
\frac{1}{2\lambda_s^{(2)}}\ln(2\cos(R_1\pi-\frac{i}{2}\oint_{K_1}Z^-(y,\lambda_s^{(2)})dy))\nn\\
q=0,1,2,...,\;r=0,1,...,[\lambda_s^{(2)}]-1
\label{631}
\ee

Finally for the limit $\lambda_s^{(3)}\equiv\lambda_{(2p+1)s+p}^{(1)}\to\infty$ we get of course again the result \mref{63}
while for $\lambda_s^{(3)}\equiv\lambda_{(2q+1)s+q}^{(2)}\to\infty$ we get:
\be
\int_{K_1(\zeta_{2,qr}^{(1)}(\lambda_s^{(3)}))}\ll(\fr\sqrt{W_{10}(y)}+
\frac{1}{2\lambda_s^{(3)}}Z^-(y,\lambda_s^{(3)})\r)dy=\nn\\
(q[\lambda_s^{(3)}]+r+(2p+1)s+p+\fr)\frac{i\pi}{\lambda_s^{(3)}}-\nn\\
\frac{1}{4\lambda_s^{(3)}}\sum_{n>k}\left(\frac{1}{2\lambda_s^{(3)}}\right)^n\ll(\oint_{K_1}+
\frac{2p+1}{2q+1}\oint_{K_2}\r)X_{2n+1}(y)dy-\nn\\
\frac{1}{2\lambda_s^{(3)}}\ln\ll(2\sin\frac{i}{2}\sum_{n>k}\left(\frac{1}{2\lambda_s^{(3)}}\right)^n\ll(\oint_{K_1}+
\frac{2p+1}{2q+1}\oint_{K_2}\r)X_{2n+1}(y)dy\r)\nn\\
q=0,1,2,...,\;r=0,1,...,[\lambda_s^{(3)}]-1
\label{632}
\ee

The last result is a modification of the formula \mref{631} when the relation \mref{5731} is taken into account.

{\it The case} $4.'$

Considering this case it is necessary to calculate the limits $\lambda_s^{(1,2,3)}\to\infty$ of $\ln E_2$.

To this goal in the case of $\lambda_s^{(1)}\to\infty$ we can use $\ln E_2=\ln(E_2/E_1)+\ln E_1$ together with \mref{371}
and \mref{61} to obtain:
\be
\frac{E_2}{E_1}=\frac{\chi_{5\to 7}\chi_{5\to{\bar 5}}}{\chi_{3\to 5}E_4}e^{2\lambda\int_{z_6}^{z_3}\sqrt{W_{10}(y)}dy-
\lambda\oint_{K_2}\sqrt{W_{10}(y)}dy}
\label{64}
\ee

And since in the limit considered
\be
E_4(\lambda_s^{(1)})\to\nn\\
\chi_{{\bar 5}\to 7}^{as}\ll[1+\exp\ll(-\lambda_s^{(1)}\oint_{K_2}\sqrt{W_{10}(y)}dy-
\sum_{n{\geq}0}\left(\frac{1}{2\lambda_s^{(1)}}\right)^{2n+1}\oint_{K_2}X_{2n+1}(y)dy\r)\r]\neq 0
\label{641}
\ee
then for the distribution of zeros of $\psi_1(z,\lambda_s^{(1)})$ in the direction of the infinite SL emerging from $z_3$ we get
to {\it all} orders in $\lambda_s^{(1)}$:
\be
\int_{K_3(\zeta_{3,qr}^{(1)}(\lambda_s^{(1)}))}\ll(\fr\sqrt{W_{10}(y)}+
\frac{1}{2\lambda_s^{(1)}}Z^-(y,\lambda_s^{(1)})\r)dy=
(q[\lambda_s^{(1)}]+r-\frac{1}{4}-\frac{R_2}{2})\frac{i\pi}{\lambda_s^{(1)}}+\nn\\
\frac{1}{4\lambda_s^{(1)}}\oint_{K_2}Z^-(y,\lambda_s^{(1)})dy+
\frac{1}{2\lambda_s^{(1)}}\ln(2\cos(R_2\pi+\frac{i}{2}\oint_{K_2}Z^-(y,\lambda_s^{(1)})dy))\nn\\
q=0,1,2,...,\;r=0,1,...,[\lambda_s^{(1)}]-1
\label{642}
\ee

The last result shows its deep difference with the formula \mref{363}, i.e. the left well quantization removes all zeros
from the infinite SL emerging from $z_2$ to shift them all on the SL emerging now from the turning point $z_3$ (see Fig.5a).

Consider now the limit $\lambda_s^{(2)}\to\infty$. In this case
\be
E_2\to\chi_{1\to{\bar 3}}^{as}(\lambda_s^{(2)})\ll(1+e^{-\oint_{K_1}\ll(\lambda_s^{(2)}\sqrt{W_{10}(y)}+
Z^-(y,\lambda_s^{(2)})\r)dy}\r)\neq 0
\label{643}
\ee
and therefore the form of the formula \mref{363} remains essentially unchanged in this limit but it is not
enough to substitute $\lambda_{s_1}$ by $\lambda_s^{(2)}$ there. As previously we have to take into
account a dependence of $R_1$ on $\lambda_s^{(2)}$ in \mref{363}. But this is the same task as
the previous one and finally we get in this case:
\be
\int_{K_2(\zeta_{2,qr}^{(1)}(\lambda_s^{(2)}))}\ll(\fr\sqrt{W_{10}(y)}+
\frac{1}{2\lambda_s^{(2)}}Z^-(y,\lambda_s^{(2)})\r)dy=(q[\lambda_s^{(2)}]+r-\frac{1}{4}-\frac{R_1}{2})\frac{i\pi}{\lambda_s^{(2)}}-\nn\\
\frac{1}{4\lambda_s^{(2)}}\oint_{K_1}Z^-(y,\lambda_s^{(2)})dy-
\frac{1}{2\lambda_s^{(2)}}\ln(2\cos(R_1\pi-\frac{i}{2}\oint_{K_1}Z^-(y,\lambda_s^{(2)})dy))\nn\\
q=0,1,2,...,\;r=0,1,...,[\lambda_s^{(2)}]-1
\label{644}
\ee
i.e. exactly the same picture of zeros ditributions around
the SL emerging from $z_2$ as in the limit $\lambda_{s_1}\to\infty$ considered in the formula \mref{363}.

It is worth to note that the formulae \mref{642} and \mref{644} are copies of each other reflecting a symmetry of the SG
considered with respect to the asymptotic quantized solutions $\psi_1^{as}(z,\lambda_s^{(1,2)})$ and
$\psi_7^{as}(z,\lambda_s^{(1,2)})$. These solutions have of course to coincide, but the solutions
$\psi_{1,7}^{as}(z,\lambda_s^{(1)})$ behave as they would be quantized in the left well only while the solutions
$\psi_{1,7}^{as}(z,\lambda_s^{(2)})$ as they would be quantized only in the right one. This observation will be
confirmed in our further calculations.

At last we have to consider the cases when
$\lambda_s^{(3)}=\lambda_{(2p+1)r+p}^{(1)}\to\infty$ and $\lambda_s^{(3)}=\lambda_{(2q+1)r+q}^{(2)}\to\infty$ when the
equalities $\lambda_{(2p+1)r+p}^{(1)}=\lambda_{(2q+1)r+q}^{(2)}$ are satisfied to $k$-th order. It easy to note that
both the formulae \mref{642} and \mref{644} have to be modified in a similar way we have got the formula \mref{632} from
\mref{5731}. Therefore we get readily:
\be
\int_{K_3(\zeta_{3,q'r}^{(1)}(\lambda_s^{(3)}))}\ll(\fr\sqrt{W_{10}(y)}+
\frac{1}{2\lambda_s^{(3)}}Z^-(y,\lambda_s^{(3)})\r)dy=\nn\\
(q'[\lambda_s^{(3)}]+r+(2q+1)s+q+\fr)\frac{i\pi}{\lambda_s^{(3)}}-\nn\\
\frac{1}{4\lambda_s^{(3)}}\sum_{n>k}\left(\frac{1}{2\lambda_s^{(3)}}\right)^n\ll(\oint_{K_2}+
\frac{2q+1}{2p+1}\oint_{K_1}\r)X_{2n+1}(y)dy-\nn\\
\frac{1}{2\lambda_s^{(3)}}\ln\ll(2\sin\frac{i}{2}\sum_{n>k}\left(\frac{1}{2\lambda_s^{(3)}}\right)^n\ll(\oint_{K_2}+
\frac{2q+1}{2p+1}\oint_{K_1}\r)X_{2n+1}(y)dy\r)\nn\\
q'=0,1,2,...,\;r=0,1,...,[\lambda_s^{(3)}]-1
\label{632}
\ee
for the first limit and
\be
\int_{K_2(\zeta_{2,q'r}^{(1)}(\lambda_s^{(3)}))}\ll(\fr\sqrt{W_{10}(y)}+
\frac{1}{2\lambda_s^{(3)}}Z^-(y,\lambda_s^{(3)})\r)dy=\nn\\
(q'[\lambda_s^{(3)}]+r+(2p+1)s+p+\fr)\frac{i\pi}{\lambda_s^{(3)}}-\nn\\
\frac{1}{4\lambda_s^{(3)}}\sum_{n>k}\left(\frac{1}{2\lambda_s^{(3)}}\right)^n\ll(\oint_{K_1}+
\frac{2p+1}{2q+1}\oint_{K_2}\r)X_{2n+1}(y)dy-\nn\\
\frac{1}{2\lambda_s^{(3)}}\ln\ll(2\sin\frac{i}{2}\sum_{n>k}\left(\frac{1}{2\lambda_s^{(3)}}\right)^n\ll(\oint_{K_1}+
\frac{2p+1}{2q+1}\oint_{K_2}\r)X_{2n+1}(y)dy\r)\nn\\
q'=0,1,2,...,\;r=0,1,...,[\lambda_s^{(3)}]-1
\label{633}
\ee
for the second one.

{\it The case} $5.'$

It follows from \mref{38} that in this case we have to consider the limit $\lambda_s^{(1,2,3)}\to\infty$ of
$\ln(E_2/E_1)$. In the cases $\lambda_s^{(2)}\to\infty$ and $\lambda_s^{(3)}=\lambda_{(2q+1)s+q}^{(2)}\to\infty$ we
can proceed directly using \mref{38} to get:
\be
\int_{K_6(\zeta_{6,q'r}^{(1)}(\lambda_s^{(2,3)}))}\ll(\fr\sqrt{W_{10}(y)}+\frac{1}{2\lambda_s^{(2,3)}}Z^-(y,\lambda_s^{(2,3)})\r)dy=
-(q'[\lambda_s^{(2,3)}]+r-\frac{1}{4})\frac{i\pi}{\lambda_s^{(2,3)}}\nn\\
q'=0,1,2,3,...,\;r=0,1,...,[\lambda_s^{(2,3)}]-1
\label{40}
\ee

In the cases $\lambda_s^{(1)}\to\infty$ and $\lambda_s^{(3)}=\lambda_{(2p+1)s+p}^{(1)}\to\infty$ we have to make use of
\mref{64}. In this way in the first case we come back to the formula \mref{642} while in the second case we get again the
formula \mref{632}.

{\it The case} $6.'$

As it follows directly from the formula \mref{41} for $\psi_1(z,\lambda)$ we have to consider
the limit $\lambda_s^{(1,2,3)}\to\infty$ of $\ln(-i{\bar A}/A)$. This limit follows however
directly from the quantization condition \mref{52} to be equal to $\ln(i\chi_{5\to 7}/\chi_{{\bar 5}\to 7})$. Therefore
independently of how $\lambda$ is quantized we have for this case:
\be
\int_{K_4(\zeta_{4,qr}^{(1)}(\lambda_s^{(i)}))}\ll(\fr\sqrt{W_{10}(y)}+\frac{1}{2\lambda_s^{(i)}}Z^-(y,\lambda_s^{(i)})\r)dy=
-(q[\lambda_s^{(i)}]+r-\frac{1}{4})\frac{i\pi}{\lambda_s^{(i)}}\nn\\
q=0,1,2,3,...,q_2,\;r=0,1,...,[\lambda_s^{(i)}]-1
\label{66}
\ee
with integer $q_2\geq 0$ given by $\int_{z_3}^{z_4}\sqrt{W_{10}(y)}dy=(q_1+r_1)i\pi,\;0\leq r_2<1$.

{\it The case} $7.'$

From the form \mref{54} of the quantization formula and the formula \mref{47} we get
readily for this case:
\be
\int_{K_7(\zeta_{7,qr}^{(1)}(\lambda_s^{(i)}))}\ll(\fr\sqrt{W_{10}(y)}+\frac{1}{2\lambda_s^{(i)}}Z^-(y,\lambda_s^{(i)})\r)dy=
-(q[\lambda_s^{(i)}]+r-\frac{1}{4})\frac{i\pi}{\lambda_s^{(i)}}\nn\\
q=0,1,2,3,...,\;r=0,1,...,[\lambda_s^{(i)}]-1
\label{67}
\ee
again independently of the way $\lambda$ is quantized.

In comparison with \mref{48} the distribution of zeros in the quantized case is now shifted totaly on the infinite SL
emerging from the turning point $z_7$.

{\it The case} $8.'$

This case is obvious since $\psi_1(z)$ coincides with $\psi_7(z)$ (up to a constant) and therefore can not have zeros
along the infinite SL emerging from $z_7$ and bounding the sector $S_7$.

We have collected the results of this analysis in Fig.5a,b for $\lambda$ quantized in the left and right well
respectively.

It follows from the figure that the limit loci of zeros of the quantized $\psi_1(z,\lambda)$ and $\psi_7(z,\lambda)$ depends on
the well in which the FS's considered are quantized in this limit. Compairing Fig.5a,b with respective Fig.4a,b we see
however that
this dependence is expressed by the way the limit zeros distributions of both the unquantized FS's
considered are unified by quantization. If these FS's are quantized in the first well the distribution of their zeros
"around" the first well is disturbed while the corresponding zeros distribution around the second well is a copy of the
unquantized $\psi_7(z,\lambda)$ and vice versa.

\section{The quantized double well - the symmetric case}

\hskip+2em The pattern of the limit loci of zeros for $\psi_1(z,\lambda)$ and $\psi_7(z,\lambda)$ quantized in the
symmetric double well
can be now easily obtained from the previous considerations as a particular unification of the two patterns of Fig.5a,b.

\begin{tabular}{c}
\psfig{figure=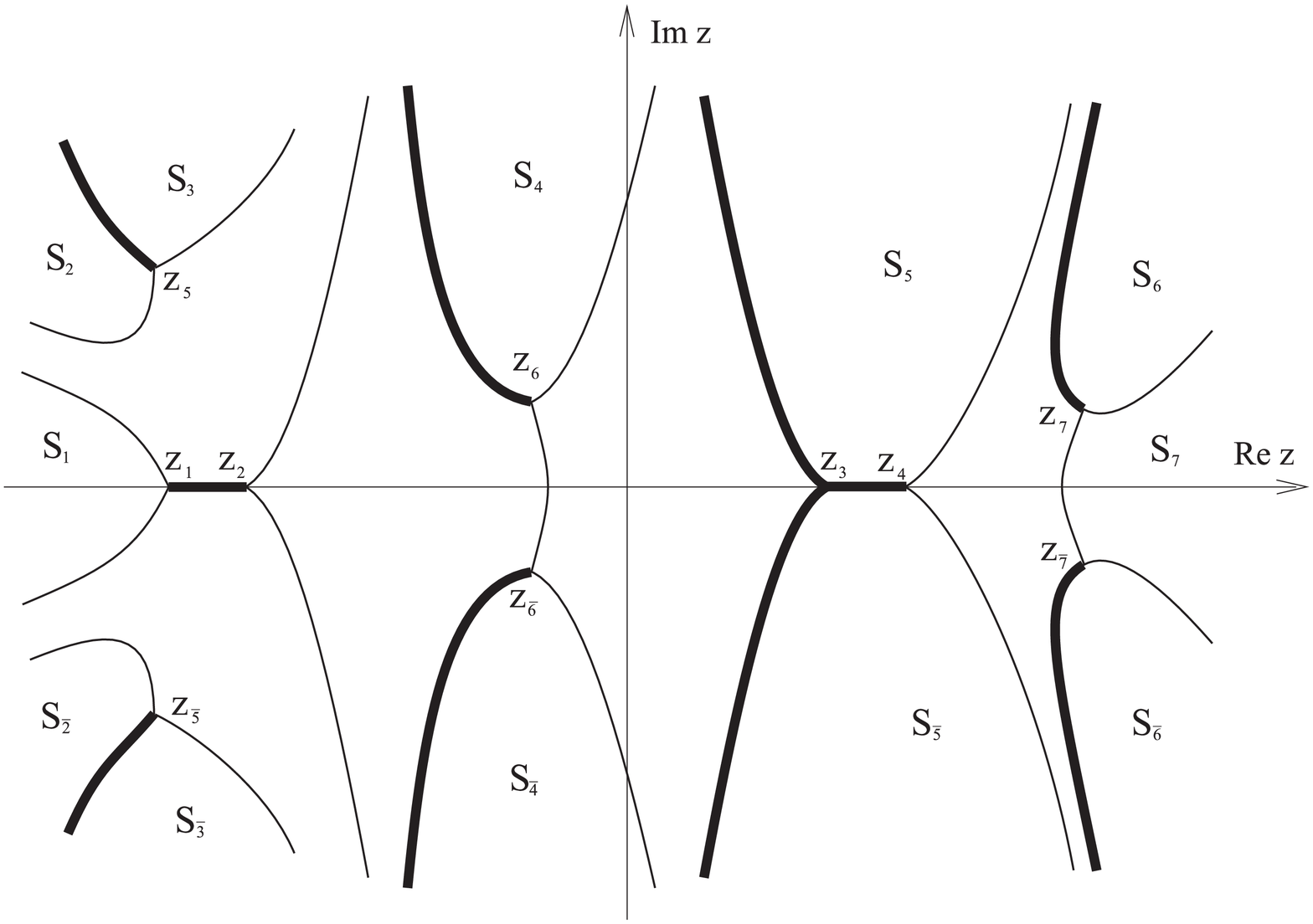,width=11cm}\\
Fig.5a ESL's (bold Stokes lines) of $\psi_1(z,\lambda)$ in the limit $\lambda_s^{(1)}\to\infty$\\
quantized in the first (left) well
\end{tabular}

\vskip 15pt

\begin{tabular}{c}
\psfig{figure=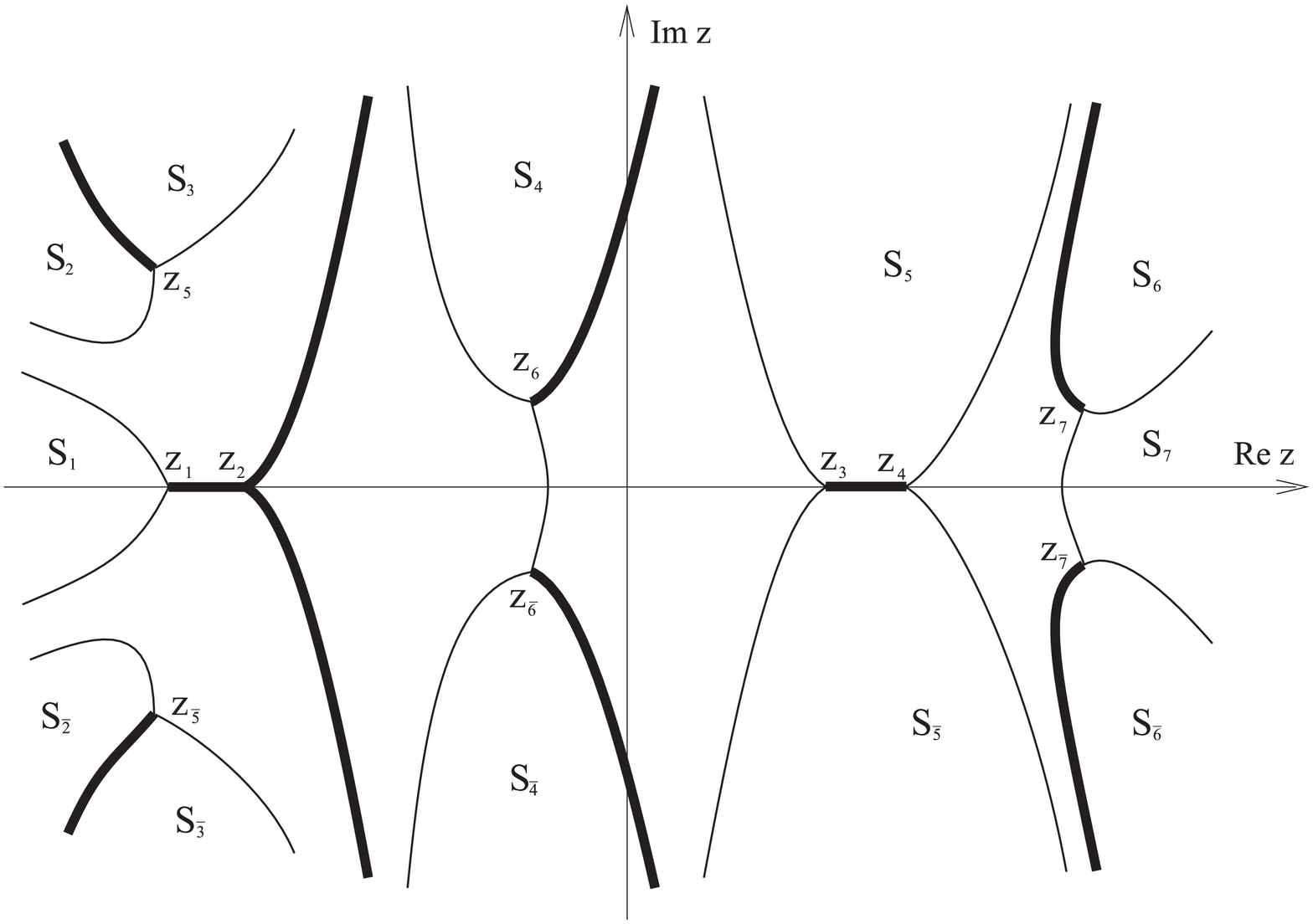,width=11cm}\\
Fig.5b ESL's (bold Stokes lines) of $\psi_1(z,\lambda)$ in the limit $\lambda_s^{(2)}\to\infty$\\
quantized in the second (right) well
\end{tabular}

\vskip 15pt

First however we have to get from the SG's of Fig.4a,b corresponding symmetric SG's. We can do in two ways. At the beginning
we put the
turning points $z_6,z_{\bar 6}$ on the imaginary axis and then apply the inversion operation to the left part of the SG
from Fig.4a or to the right one. We get in this way two cases of the symmetric double wells shown in Fig.6a,b

If both the cases are not quantized the limit zeros distributions of $\psi_1(z,\lambda)$ are shown in Fig.7a and Fig.7b
for the corresponding symmetric double-wells. For $\psi_7(z,\lambda)$ the corresponding pictures are the mirror
reflections in the imaginary axis of the last figures.

\vskip 25pt

\begin{tabular}{c}
\psfig{figure=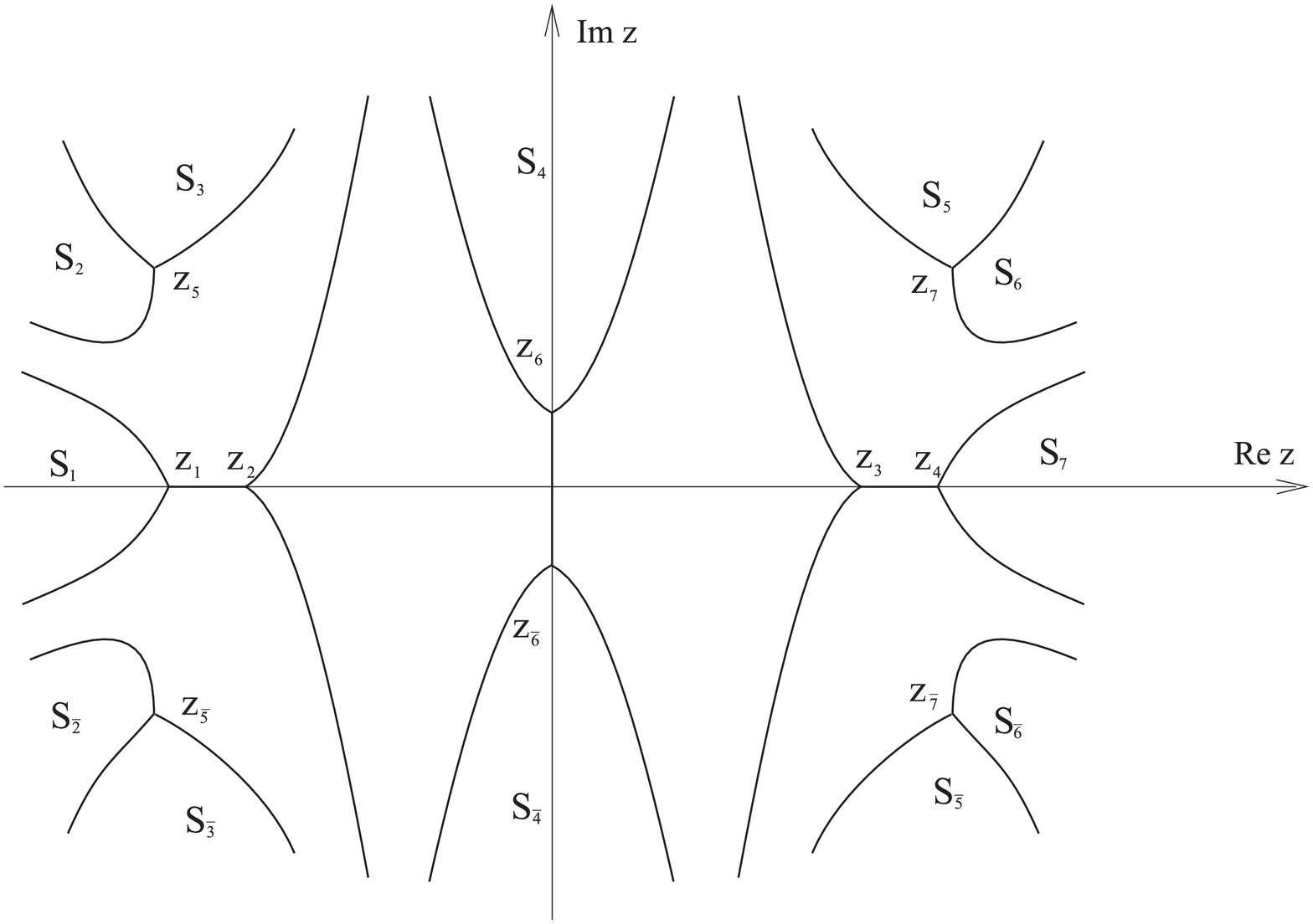,width=12cm}\\
Fig.6a The first variant of the symmetric double-well
\end{tabular}

\vskip 25pt

\begin{tabular}{c}
\psfig{figure=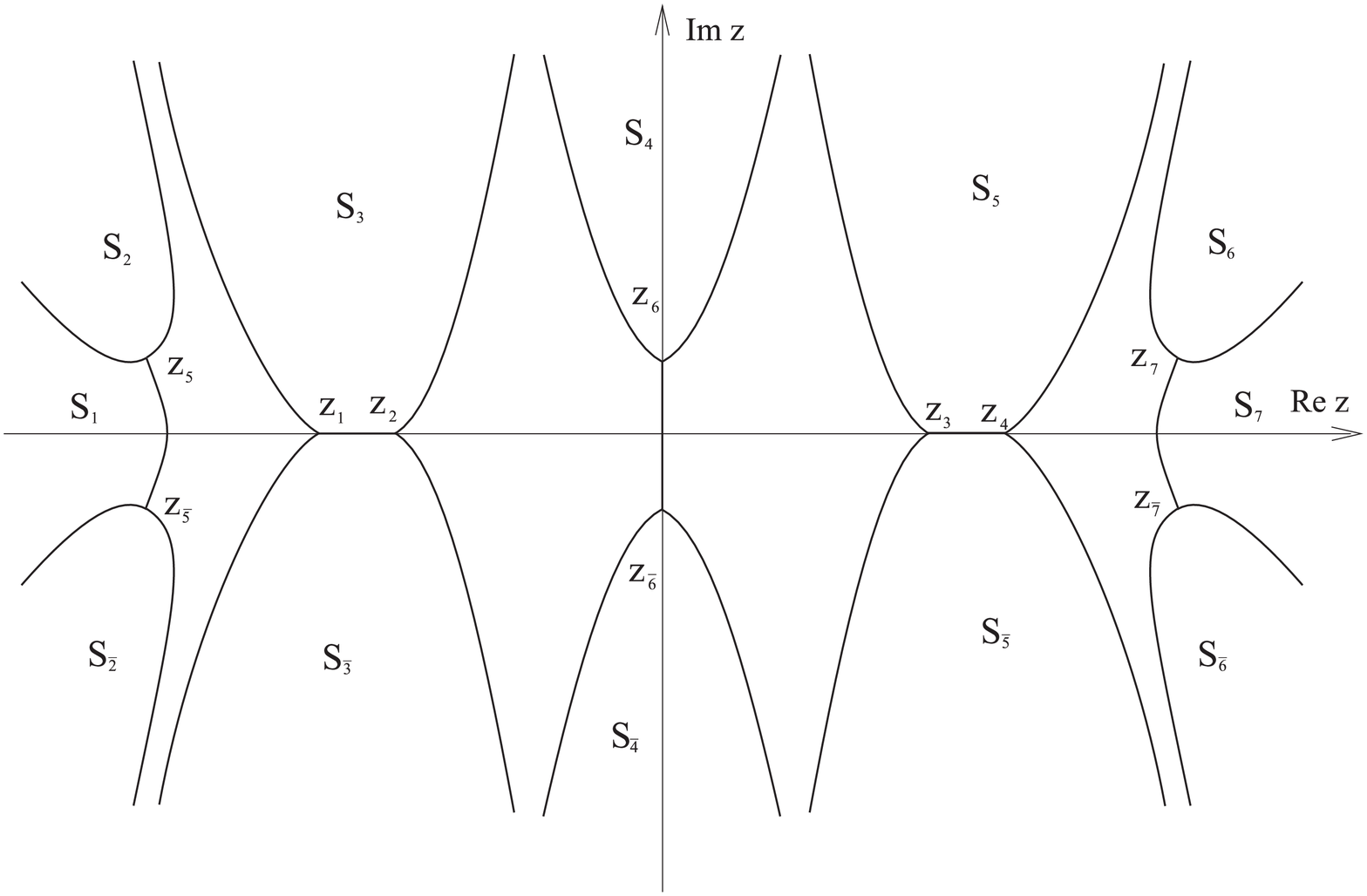,width=12cm}\\
Fig.6b The second variant of the symmetric double-well
\end{tabular}

\vskip 25pt

For the quantized cases Fig.7a and Fig.8a have to be unified with their mirror reflections just mentioned to form figures
shown in Fig.8a,b respectievly.

\vskip 25pt

\begin{tabular}{c}
\psfig{figure=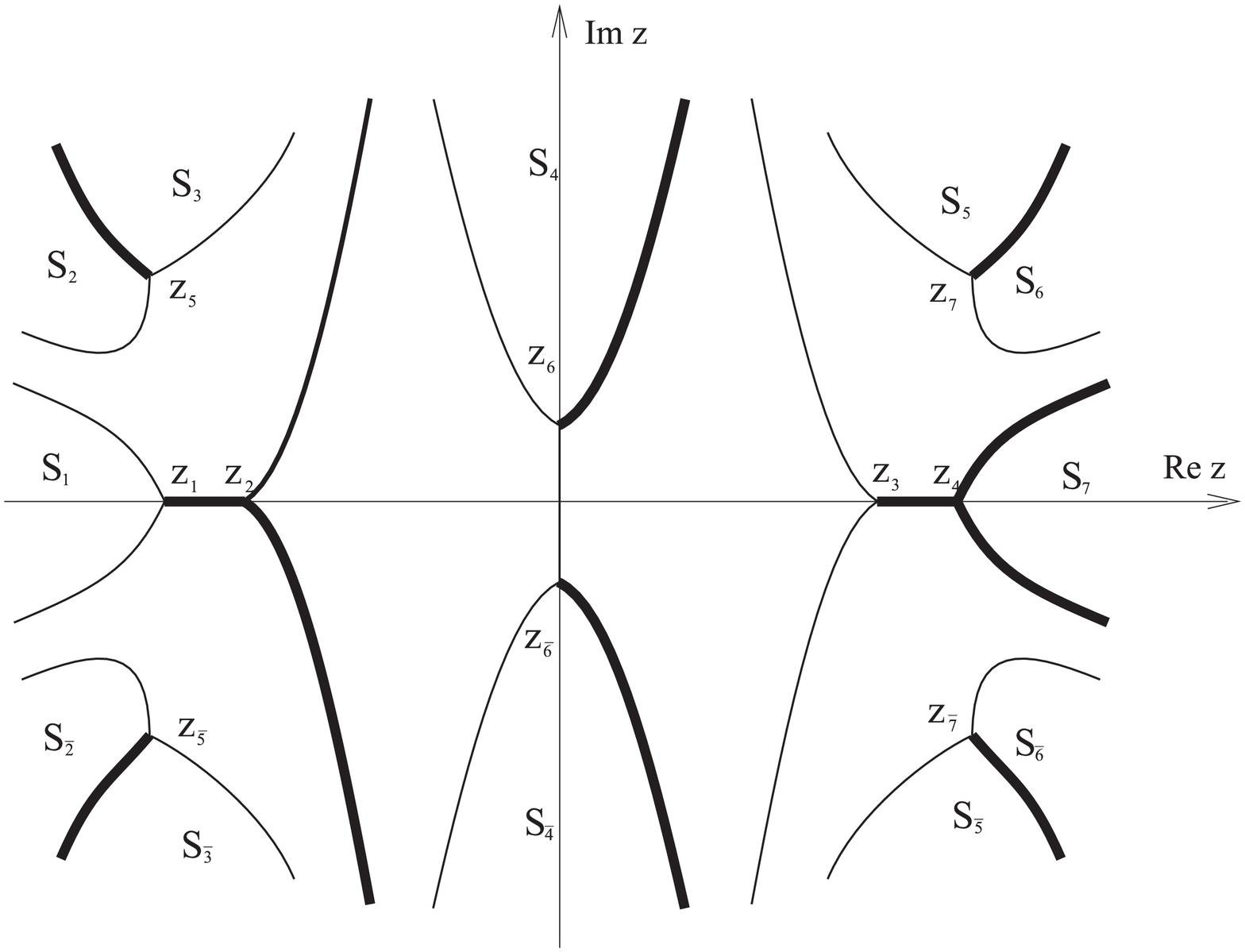,width=13cm}\\
Fig.7a ESL's (bold Stokes lines) of $\psi_1(z,\lambda)$ in the regular limits $\lambda\to\infty$.\\
The not quantized case of the first variant of the symmetric double-well
\end{tabular}

\vskip 25pt

\begin{tabular}{c}
\psfig{figure=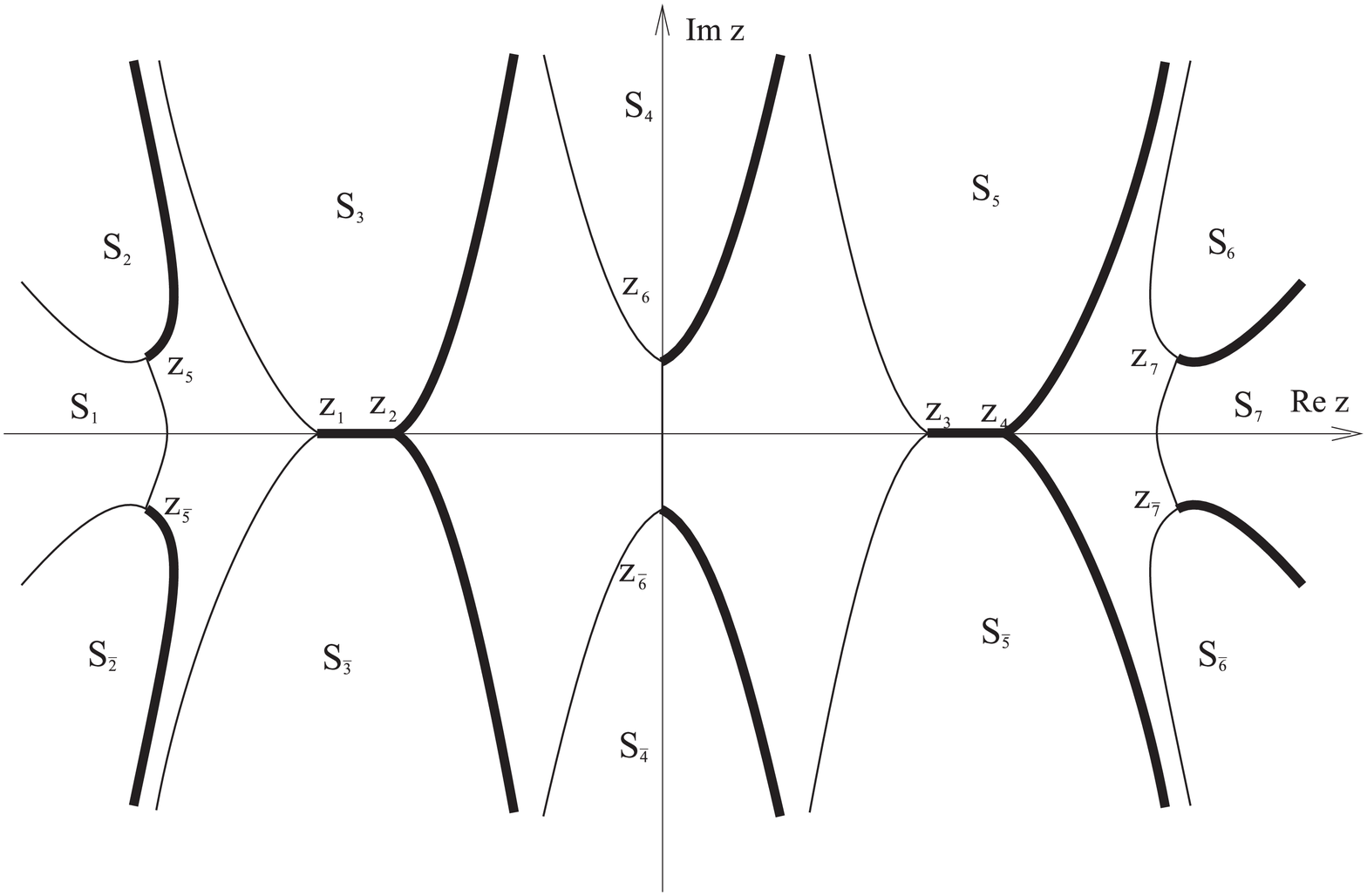,width=13cm}\\
Fig.7b ESL's (bold Stokes lines) of $\psi_1(z,\lambda)$ in the regular limits $\lambda\to\infty$.\\
The not quantized case of the second variant of the symmetric double-well
\end{tabular}

\begin{tabular}{c}
\psfig{figure=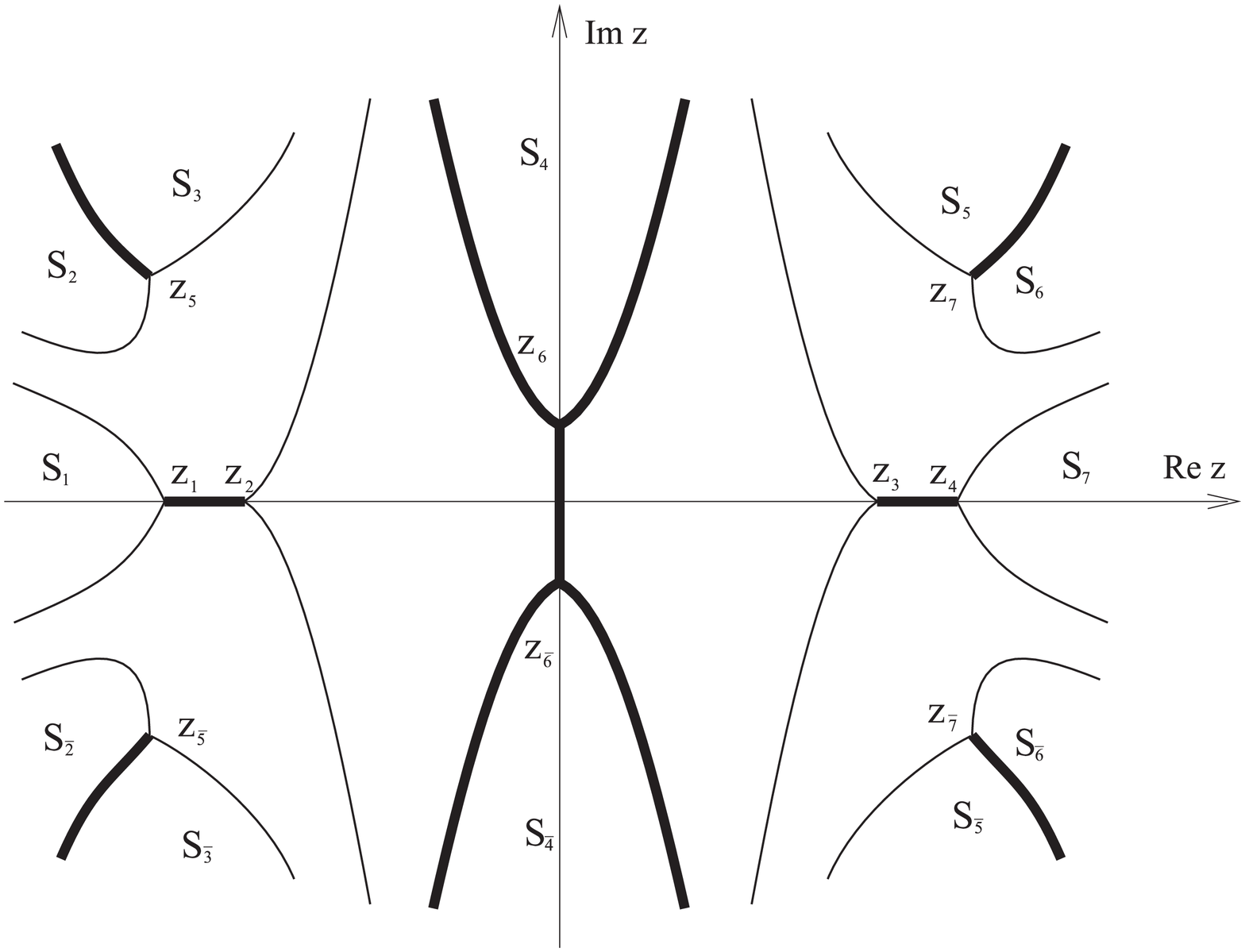,width=11cm}\\
Fig.8a ESL's (bold Stokes lines) of $\psi_1(z,\lambda_s)$ in the regular limits $\lambda_s\to\infty$.\\
The quantized case of the first variant of the symmetric double-well
\end{tabular}

\begin{tabular}{c}
\psfig{figure=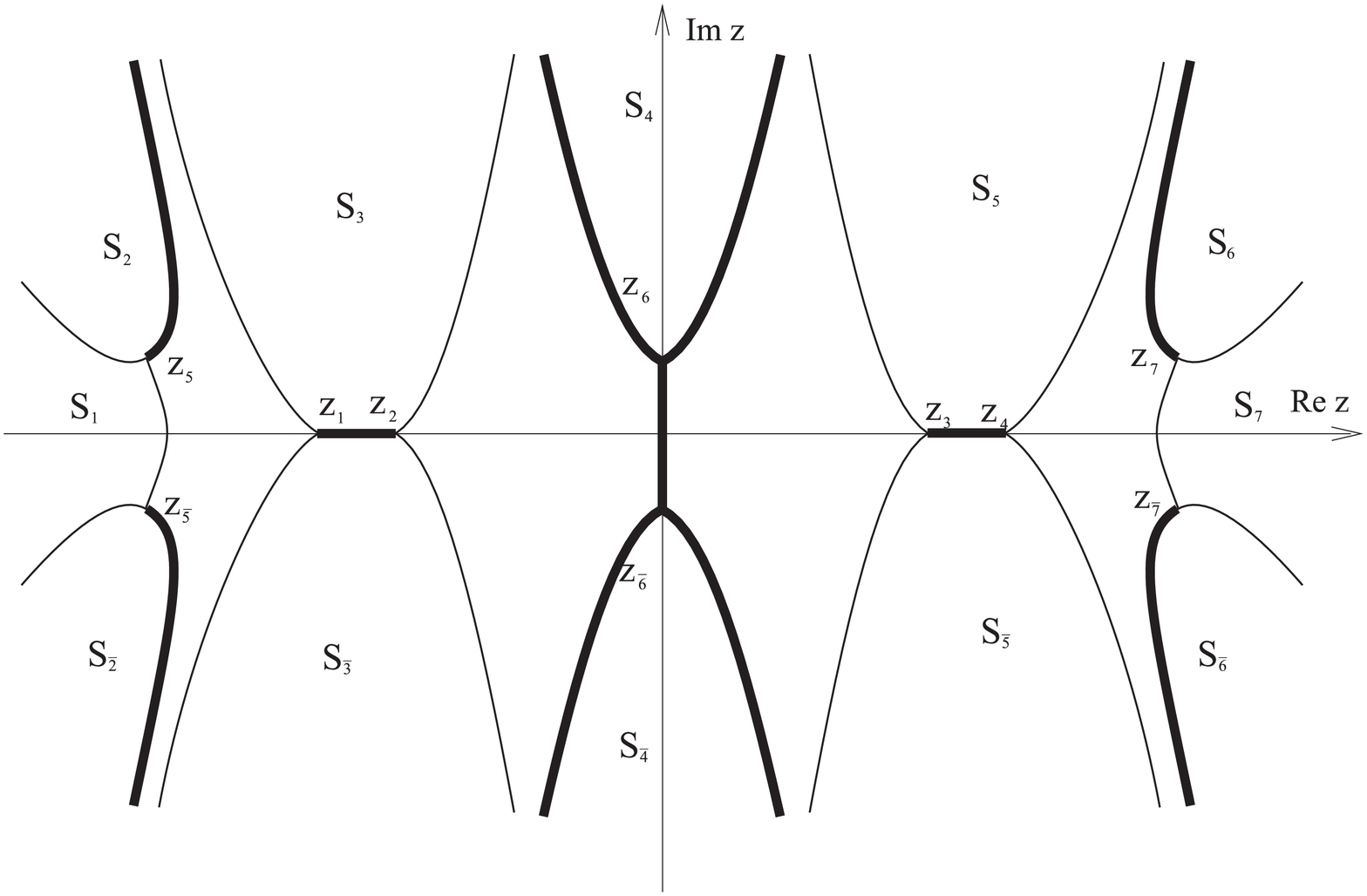,width=11cm}\\
Fig.8b   ESL's (bold Stokes lines) of $\psi_1(z,\lambda_s)$ in the regular limits $\lambda_s\to\infty$.\\
The quantized case of the second variant of the symmetric double-well
\end{tabular}

\section{Simple generalizations -- multiple-well potentials}

\hskip+2em The results obtained in the previous section suggest simple generalizations of them by enlarging a
number of wells on the real axis so the number of internal SL's lying on this axis while the remaining internal SL's
have to cross the real axis only.

However as we have mentioned in sec.4 there are no obvious ways of generalizations of the
results obtained so far when $\hbar$ is to be quantized and we have to limit ourselves rather to situations when the
quantization of $\hbar$ for small
values of it are possible and can be performed effectively. An enlarging a number of wells in the way described above
seems to satisfy these conditions.

The figures Fig.9 - Fig.12c illustrate this situation and show the limit distribution
of zeros of the FS's $\psi_1(z,\lambda)$ and $\psi_{n+2}(z,\lambda)$ in all basic variants of mutual relations between
these solutions.

To get these figures we have applied the rules given below which can be read off from the
results of the previous sections.

For this goal let us remind a general observation mentioned in sec.5
which generates the rules. This is that non-critical SG's arise in the case of a fixed polynomial potential from
critical ones by small changes of $\arg\lambda$ so that three SL's emerging from each root of the polynomial rotate around
the root. During these rotations continuously glued internal and external SL's split into external ones only leading to
non-critical SG's. Since however in the latter cases the limit locus of zeros of FS's are known to lie on their ESL's
({\bf Theorem 1a}) then by the continuity argument these limit zeros loci of the considered FS have had to coincide
earlier with the splitting
SL's of the critical case. For the non-quantized critical SG's this argument allows us to identify ESL's of such critical
SG's immediately. For the solutions $\psi_1(z,\lambda)$ and $\psi_{n+2}(z,\lambda)$ it leads to the patterns of Fig.Fig.
9-10.

However when the solutions $\psi_1(z,\lambda)$ and $\psi_{n+2}(z,\lambda)$ are matched according to the quantization condition
defined by the $q^{th}$-well (see Fig.11) then to the four external SL's emerging from the two turning points defining
the $q^{th}$-well applies {\bf Corollary 1},
i.e. these SL's are not any longer ESL's of any of these solutions while the internal SL linking these two turning points
of the well still maintains to be ESL.

All the other ESL's however satisfy the
rules described above for the non quantized case independently for each solution with the restriction that to the left
from the $q^{th}$-well the ESL's are determined by $\psi_1(z,\lambda)$ while to the right from the well the system of
ESL's is determind by $\psi_{n+2}(z,\lambda)$. The ESL's which lie inside the $q^{th}$-well are determind according to the above
rules by both the solutions simultanuously.

In the quantized cases ($\psi_1(z,\lambda_s)$ matched with $\psi_{n+2}(z,\lambda_s)$) for the symmetric multiple-well
potential the corresponding rules change
accordingly further collecting common features of the quantized asymmetric cases and being the following
\begin{enumerate}
\item the two internal SL's of the two symmetric quantized wells remain to be exceptional while the infinite SL's emerging
from their ends are not ESL's any longer
\item the exceptional SL's on the left from the left quantized well coincide with the exceptional ones of the not quantized
$\psi_1(z,\lambda_s)$, these on the right from the right quantized well with the exceptional lines of the not quantized
$\psi_{n+2}(z,\lambda_s)$ while those emerging from the complex turning points occupying the two quantized wells are
unchanged (in comparison with the unquantized case)
\item the exceptional SL's between the left quantized well and the vertical symmetry axis of the SG considered remain the same as
for the unquantized $\psi_{n+2}(z,\lambda_s)$ while those between this axis and the right quantized well coincide with
the exceptional lines for the unquantized $\psi_1(z,\lambda_s)$
\item the middle sectors (if they are) collect all the properties of the last point
\item the exceptional lines for $\psi_1(z,\lambda_s)$ when the corresponding SG contains the single middle well and the
latter is quantized are identified by the rules described in the first two points above
\end{enumerate}

\vskip 15pt

\begin{tabular}{c}
\psfig{figure=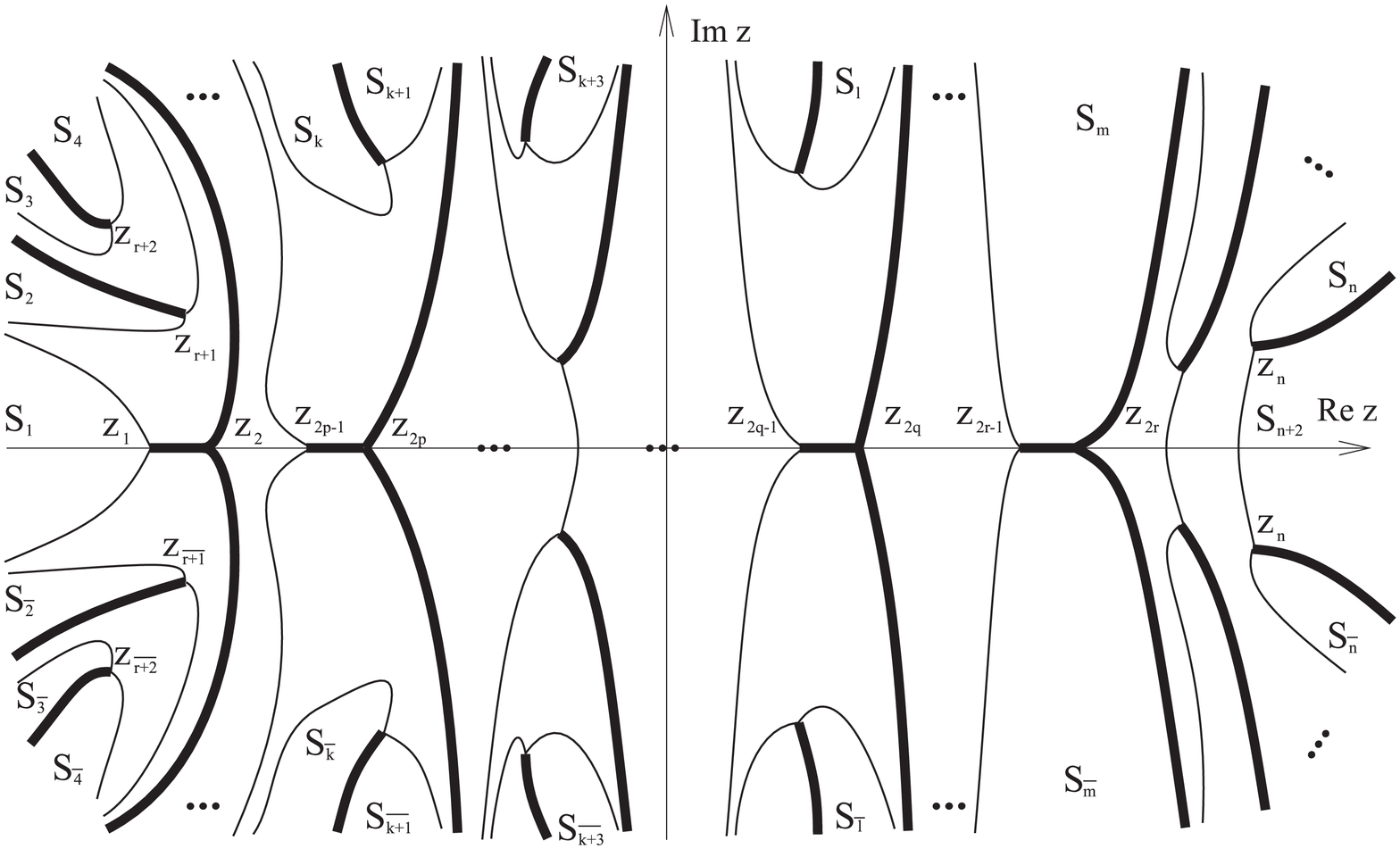,width=13cm}\\
Fig.9 ESL's (bold Stokes lines) of $\psi_1(z,\lambda_s)$ in the regular limits\\$\lambda_s\to\infty$
for $W_{2n}(z)$ polynomial potential. The non-quantized case
\end{tabular}

\vskip 15pt

\begin{tabular}{c}
\psfig{figure=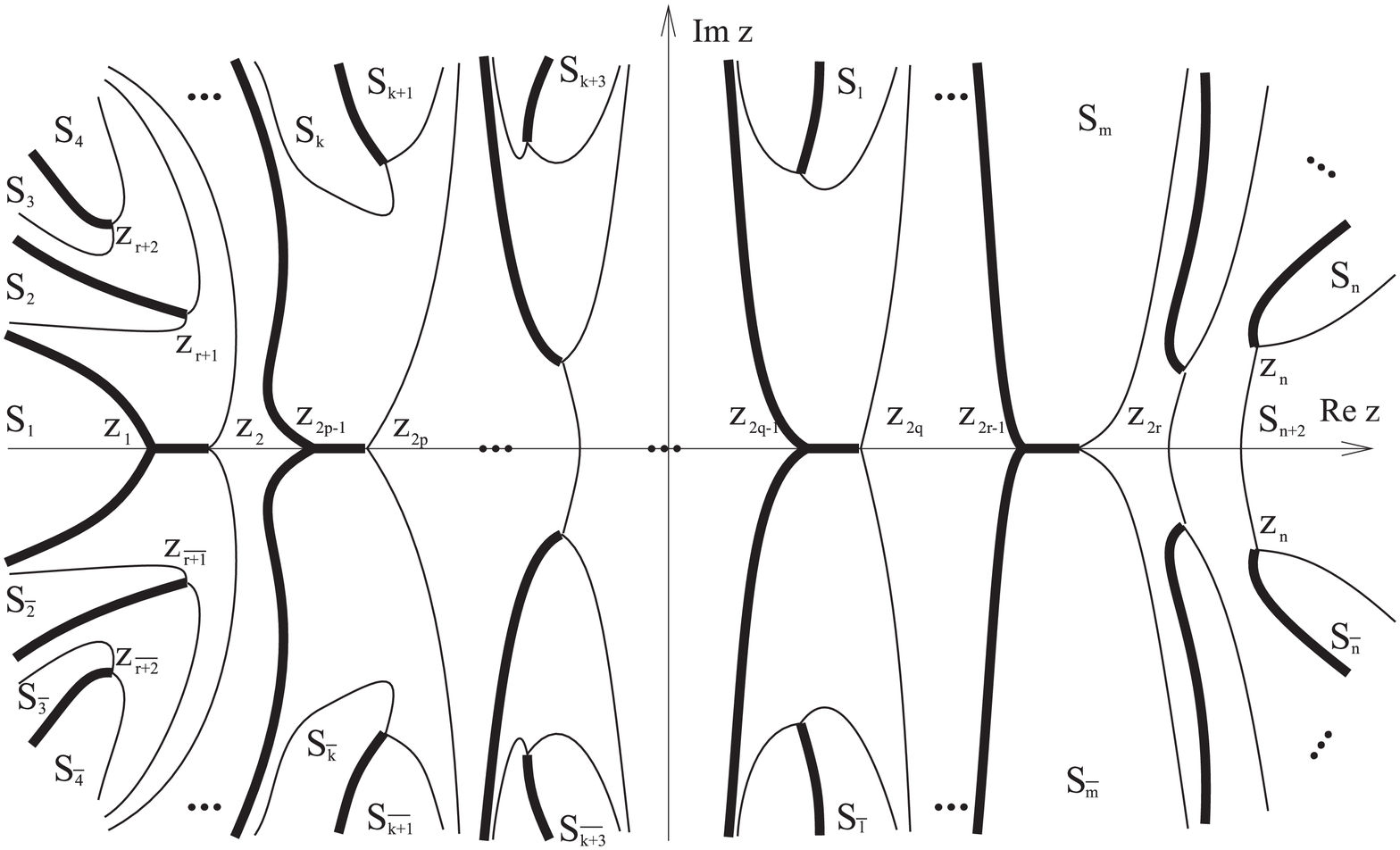,width=13cm}\\
Fig.10 ESL's (bold Stokes lines) of $\psi_{n+2}(z,\lambda_s)$ in the regular limits\\$\lambda_s\to\infty$
The non-quantized case
\end{tabular}

\begin{tabular}{c}
\psfig{figure=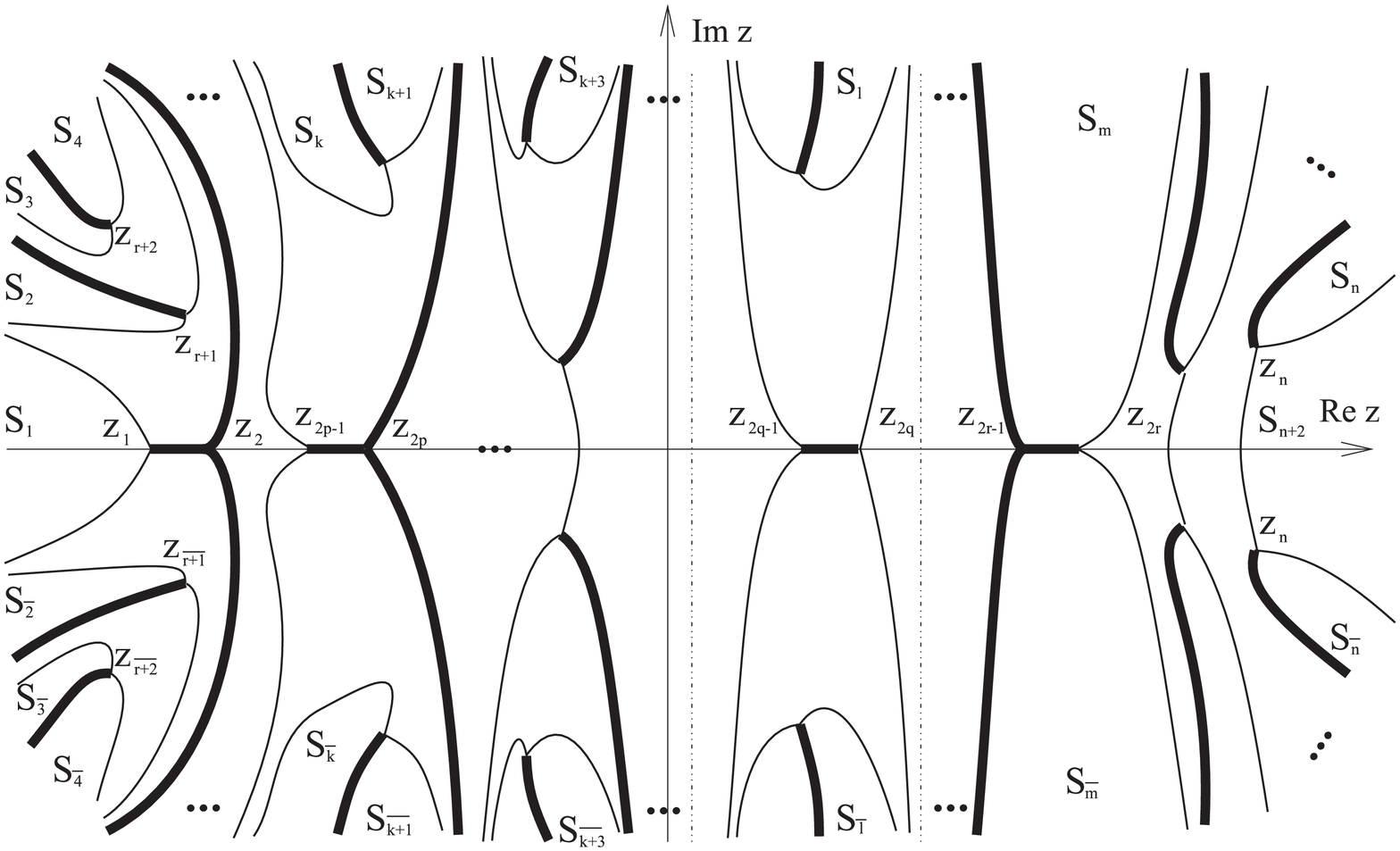,width=14cm}\\
Fig.11 ESL's (bold Stokes lines) of $\psi_1(z,\lambda_q)$=$C\psi_{n+2}(z,\lambda_q)$ in the regular\\
limit $\lambda_q\to\infty$ (bold Stokes lines). $\lambda_q$ is quantized in the $q^{th}$-well
\end{tabular}

\vskip 18pt

\begin{tabular}{c}
\psfig{figure=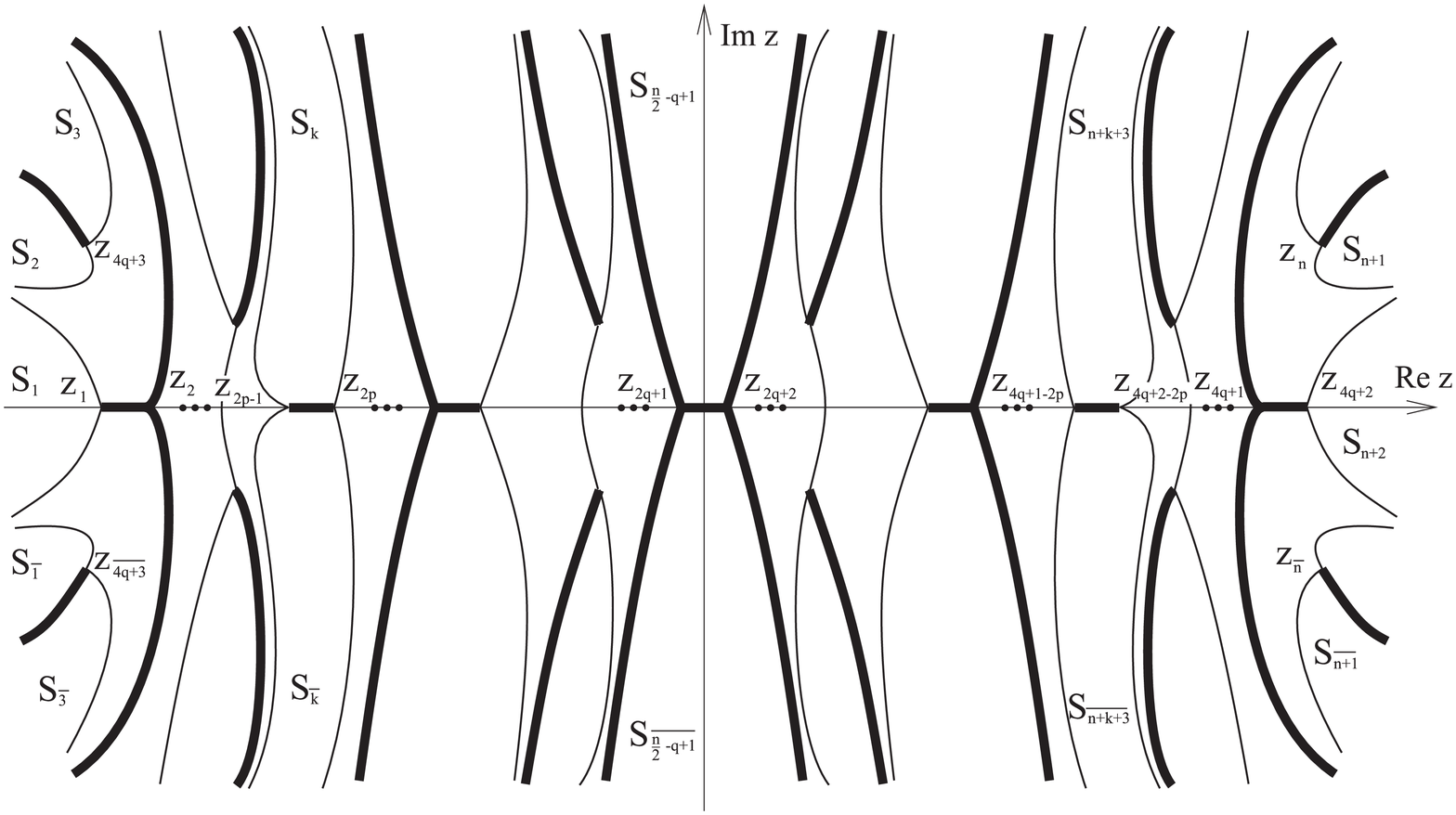,width=14cm}\\
Fig.12a ESL's (bold Stokes lines) of $\psi_1(z,\lambda_s^{(p)})=$$C\psi_{n+2}(z,\lambda_s^{(p)})$ in\\the regular
limit $\lambda_s^{(p)}\to\infty$. The symmetric multiple-well\\variant quantized in the $p^{th}$-well
\end{tabular}

\vskip 18pt

As we have mentioned the rules formulated above can be easily extracted from the previous sections but they can be also
proved by the same
methods as used previously. In particular all the ways of obtaining the limit $\lambda\to\infty$ considered earlier and
the limit loci of zeros induced by them are applied also in these generalizations so that the corresponding formulae
of the previous sections can be rephrased accordingly with a respective effort.

\vskip 18pt

\begin{tabular}{c}
\psfig{figure=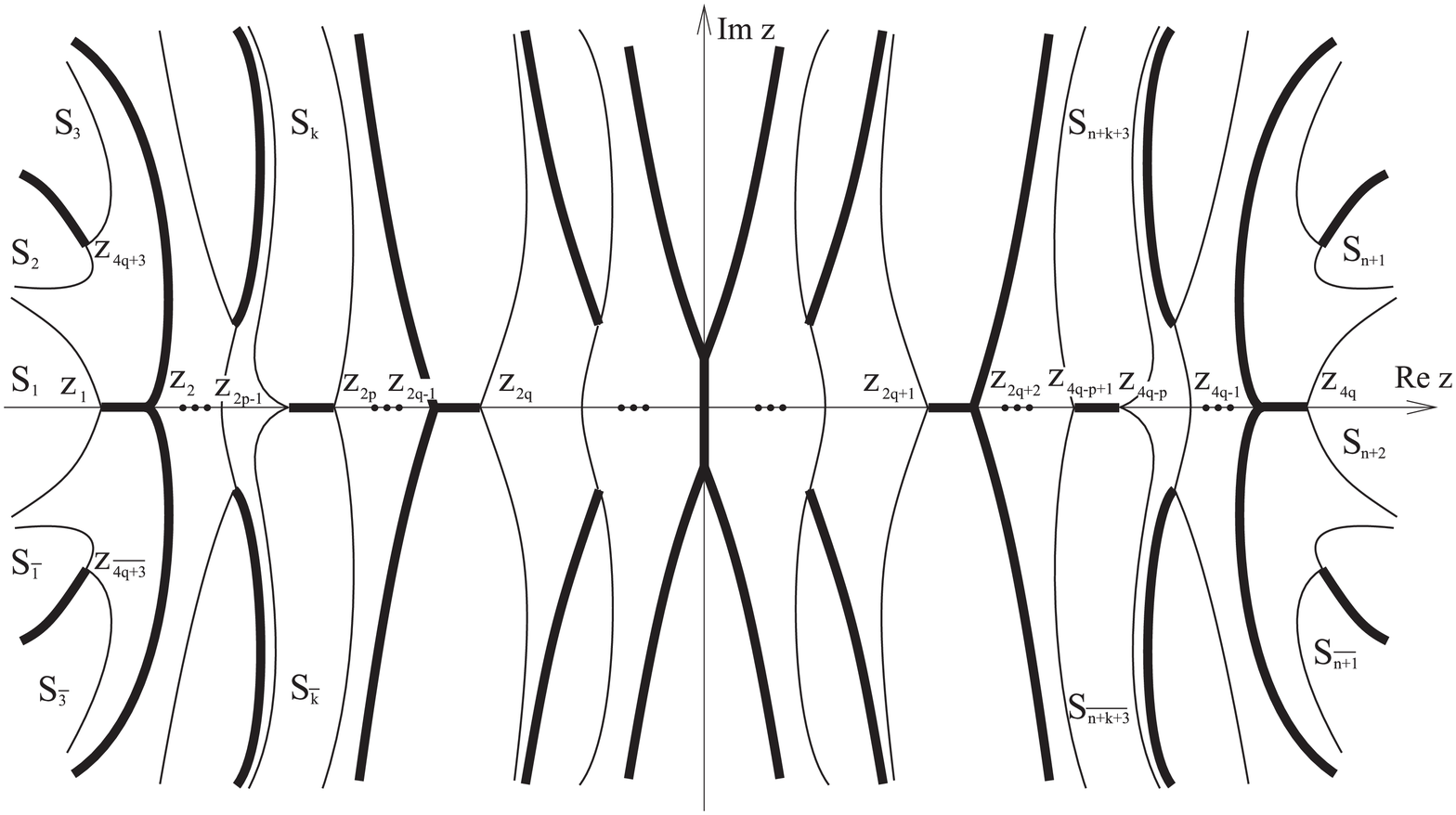,width=14cm}\\
Fig.12b ESL's (bold Stokes lines) of $\psi_1(z,\lambda_s^{(p)})=$$C\psi_{n+2}(z,\lambda_s^{(p)})$ in\\the regular
limit $\lambda_s^{(p)}\to\infty$. The second symmetric multiple-well\\variant quantized in the $p^{th}$-well
\end{tabular}

\vskip 18pt

\begin{tabular}{c}
\psfig{figure=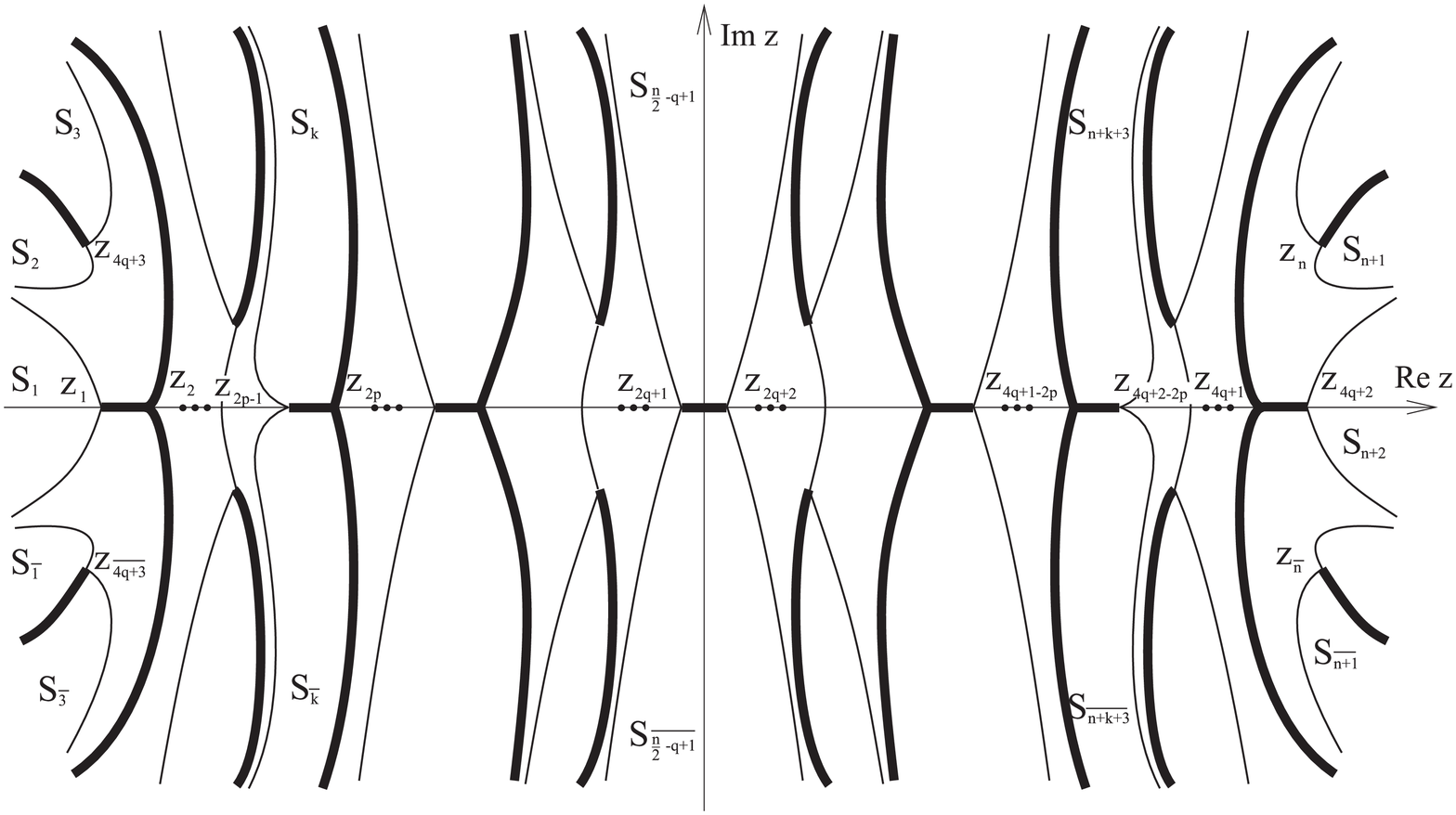,width=14cm}\\
Fig.12c  The regular limit $\lambda_s^{(q+1)}\to\infty$ of loci of zeros of $\psi_1(z,\lambda_s^{(q+1)})=$\\
$C\psi_{n+2}(z,\lambda_s^{(q+1)})$ (bold Stokes lines). The symmetric multiple-well case\\quantized in the middle well
\end{tabular}

\section{Summary and discussion}

\hskip+2em In this paper we have considered the small-$\hbar$ semiclassical limit for a general non-critical case of SG
with simple
turning points for which we have got the general {\bf Theorem 1a}. For the critical case with a unique inner SL we have got
{\bf Theorem 1b}. It appeared that both the last theorems looked very similarly to the corresponding {\bf Teorems 2a,b} of
our previous paper \cite{11} when the high energy semiclassical limit was considered. The quantized case however needs some
special conditions to be satisfied for the case to be considered.

In the case of SG's with more than a unique inner SL a pattern of the limit zeros loci of fundamental solutions has appeared
to be not essentially different in comparison with the previous cases of SG's considered. While we have limited ourselves
initially
to a real D-W potential of tenth-degree its degree has appeared in fact not much important, i.e. we have been able to
formulate a generalization of this case by enlarging a number of wells and enhancing arbitralily the potential degree.
Nevertheless we have concluded that when the $\hbar$-variable eigenvalue problem was considered for a polynomial potential
then a special arrangement of roots of the potential was necessary as well as the special choice of accompanying FS's
to provide the problem for which the small-$\hbar$ semiclassical limit exists.

Our calculations are in agreement with those of Hezari's results \cite{9} the latter being particular cases of ours. In the language of Hezari's paper we can express our results in the double-well case, both symmetric or
not, that there are a number of zero limit measures defined on SL's shown in Fig. 7-13 and given by the integrals of the
formulae \mref{2721} and similar others. It is clear that in the multiple-well cases a number of zero limit measures grows
accordingly to a number of wells and to relations between the phase integrals corresponding to the wells. Nevertheless
a general picture of zeros distributions along the Stokes lines is such as described by the rules formulated in the
sections 3 and 7.

\section*{Acknowledgment}

\hskip+2em I would like to thank to dr Haim Hezari for his interest to my paper and for his pointing me out that there
are no inconsistences in his callculations made in his paper \cite{9} what I erroneuosly suggested to be in the
previous versions of my paper.

\section*{Appendix A}

\hskip+2em The formula \mref{14a} has the following asymptotic expansion allowing us to calculate the semiclassical
expansion \mref{141}:
\be
\int_{K_l(\zeta_{l,qr0}^{(k)}(\Lambda))}\ll(\fr\sqrt{W_n(y)}-
\frac{1}{2\lambda}Z_k(y,\lambda)\r)dy+\nn\\
2\sum_{s\geq 1}\frac{1}{s!}\ll.\ll(\fr\sqrt{W_n(y)}-
\frac{1}{2\lambda}Z_k(y,\lambda)\r)^{(s)}\r|_{y=\zeta_{l,qr0}^{(k)}(\Lambda)}
\ll(\sum_{p\geq 1}\frac{1}{\lambda^p}\zeta_{l,qrp}^{(k)}(\Lambda)\r)^s=\nn\\
\ll(q[|\lambda|]+r-\frac{1}{4}\r)\frac{i\pi}{\lambda},\;\;\;q>0
\label{A234}
\ee
and with a similar formula for $q=0$ obtained from the last one where $\zeta_{l,qr0}^{(0)}(\Lambda)$ is substituted by
$z_l+\zeta_{l,0r1}^{(k)}(\Lambda)/\lambda$ and $p>1$.

\section*{Appendix B}

\hskip+2em We describe here the way the coefficients $\alpha_{\frac{k}{i}\to j}$ and
$\alpha_{\frac{k}{j}\to i}$ in the linear combination:
\be
\psi_k(z)=\alpha_{\frac{k}{i}\to j}\psi_i(z)+\alpha_{\frac{k}{j}\to i}\psi_j(z)
\label{B1}
\ee
are calculated.

First it is assumed that the three FS's which are involved in such a linear combination can be continued along canonical paths
to the sectors in which their partner FS's are defined, i.e. the sectors $S_x,\;x=i,j,k,$ have to communicate canonically
with themselves.

Next according to our earlier conventions for each $\psi_k(z)$ we choose as $z_k$ in the formula
\mref{10} one of TP's lying on the boundary of the corresponding sector $S_k$. Then the first coefficient in the formula
\mref{B1} can be calculated according to the rule
\be
\alpha_{\frac{k}{i}\to j}=\lim_{z\to\infty_j}\frac{\psi_k(z)}{\psi_i(z)}
\label{B2}
\ee
with the analogous formula for the second one.

Since the above rules need all $\psi$'s to be continued analytically along canonical paths on the $C_{cut}$-plane we have
to formulate rules for crossing the cuts by $\psi$'s. On the $C_{cut}$-plane there are cuts, each the same for the three
factors: $\sqrt{W_n(z)}$, $W_n^{-\frac{1}{4}}(z)$ and $\chi(z)$ of $\psi(z)$, emerging from each TP $z_k,\;k=1,...,n$.
We adopt the following
convention for calculation of the argument $\arg(z-z_k)$: 1. it is calculated with respect to the axis emerging from $z_k$ and being parallel
to the real axis; 2. it is calculated with the positive sign in the anticlockwise direction and with the negative one in the
opposite direction.

According to these conventions $\arg(z-z_k)$ jumps by $\pm\pi$ when the cut emerging from $z_k$ is crossed clockwise or
anticlockwise respectively. It then follows that $\sqrt{W_n(z)}$ changes its sign only being continued by the cut while
$W_n^{-\frac{1}{4}}(z)$ has to be multiplied by $\pm i$ when continued through the cut clockwise (with respect to $z_k$)
or anticlockwise respectively.

A corresponding change of the $\chi$-factor is a little bit more complicated, namely, the $n^{th}$-term of
the expansion \mref{11} defining it changes its sign by $(-1)^n$ when $\chi(z)$ is continued through the cut so changing
effectively it signature $\sigma$ to the opposite one, i.e. such a crossing switches the continued FS between the two
forms it has in the two sectors which are the mutuall projections of each other. This feature of FS's has been
discussed earlier in sec.2 (see \mref{13a} and the discussion following it).

The above conventions define now completely the procedure of calculating both the coefficients in the
formula \mref{B1}. Consider for example the way the formula \mref{271} has been obtained. The corresponding cuts emerging
from the TP's $z_1$ and $z_5$ could be drawn on Fig.3 to the left parallely to the real axis. We then have put:
\be
\psi_1(z,\lambda)= W_{10}^{-\frac{1}{4}}(z)e^{\lambda\int_{z_1}^{z}\sqrt{W_{10}(y)}dy}\chi_1(z,\lambda)
\label{B31}
\ee
where $z\in S_1$ and is above the cut emerging from $z_1$ so that $\sigma_1=1$.
\be
\psi_2(z,\lambda)= W_{10}^{-\frac{1}{4}}(z)e^{-\lambda\int_{z_5}^{z}\sqrt{W_{10}(y)}dy}\chi_2(z,\lambda)
\label{B32}
\ee
for $z\in S_2$ and below the cut emerging from $z_5$ so that $\sigma_2=-1$ and
\be
\psi_3(z,\lambda)= W_{10}^{-\frac{1}{4}}(z)e^{-\lambda\int_{z_5}^{z}\sqrt{W_{10}(y)}dy}\chi_3(z,\lambda)
\label{B3}
\ee
assuming $z\in S_3$ and above the cut emerging from $z_5$ so that $\sigma_{3}=-1$.

Continuing now the solutions $\psi_i(z,\lambda),\;i=1,2,$ to the sector $S_3$ along canonical paths we get :
\be
\alpha_{\frac{1}{2}\to 3}=\lim_{z\to\infty_3}\frac{\psi_1(z,\lambda)}{\psi_2(z,\lambda)}=
\lim_{z\to\infty_3}\frac{W_{10}^{-\frac{1}{4}}(z)e^{\ll(\lambda\int_{z_1}^{z_5}+\lambda\int_{z_5}^{z}\r)
\sqrt{W_{10}(y)}dy}\chi_1(z,\lambda)}
{iW_{10}^{-\frac{1}{4}}(z)e^{\lambda\int_{z_5}^{z}\sqrt{W_{10}(y)}dy}\chi_2(z,\lambda)}=\nn\\
-ie^{\lambda\int_{z_1}^{z_5}\sqrt{W_{10}(y)}dy}\chi_{1\to 3}(\lambda)
\label{B4}
\ee
where $\chi_{1\to 3}(\lambda)=\lim_{z\to\infty_3}\chi_1(z,\lambda)$ and  $\chi_{2\to 3}(\lambda)\equiv 1$.

Similarly for the second coefficient in \mref{271} we get
\be
\alpha_{\frac{1}{3}\to 2}=ie^{\lambda\int_{z_1}^{z_5}\sqrt{W_{10}(y)}dy}
\label{B5}
\ee
since now $\chi_{1\to 2}(\lambda)=\chi_{3\to 2}(\lambda)\equiv 1$.

Taking into account \mref{B4}-\mref{B5} we get \mref{271} for the case when $z\in S_3$ (the factor
$W_{10}^{-\frac{1}{4}}(z)$ of $\psi_2(z,\lambda)$ gets then additionally the factor $i$ crossing the cut emerging from
$z_5$).

In the calculations made in this paper we have made use of the relation $\chi_{i\to j}(\lambda)=\chi_{j\to i}(\lambda)$
valid for any pair of such coefficients (see for example Ref.2 of \cite{4}) as well as of the following identity which is
valid for every four different FS's $\psi_x(z,\lambda),\;x=i,j,k,l,$ which sectors they are defined in can communicate
canonically by pairs (see for example Ref. 2 of \cite{5}):
\be
\alpha_{\frac{i}{j}\to k}=\alpha_{\frac{i}{j}\to l}+\alpha_{\frac{i}{l}\to j}\alpha_{\frac{l}{j}\to k}
\label{B6}
\ee

\section*{Appendix C}

\hskip+2em We discuss here a necessity of using a pair of FS's with different signatures in forming a linear combinations
of the solution which the limit zero loci is investigated to obtain non trivial conditions for these loci.

For this goal consider the linear combination \mref{271} of
$\psi_1(z,\lambda)$ by $\psi_2(z,\lambda)$ and $\psi_3(z,\lambda)$. This combination can be continued along canonical paths
to a vicinity of the external SL emerging from $z_2$ and running to the upper infinity of the $C_{cut}$-plane. This
combination can be written there as follows
\be
\psi_1(z,\lambda)=-iW_{10}^{-\frac{1}{4}}(z)e^{\lambda\int_{z_1}^{z_5}\sqrt{W_{10}(y)}dy}\ll(\chi_{1\to 3}(\lambda)\chi_2(z,\lambda)
         -\chi_3(z,\lambda)\r)e^{-\lambda\int_{z_5}^{z}\sqrt{W_{10}(y)}dy}
\label{C1}
\ee

A condition for $\psi_1(z,\lambda)$ to vanish there is therefore
\be
\chi_{1\to 3}(\lambda)\chi_2(z,\lambda)-\chi_3(z,\lambda)=0
\label{C2}
\ee
where $z$ is close to the SL mentioned.

If however we now take in \mref{C2} the limit $\lambda\to\infty$ then with the help of the exponential representions
\mref{551}-\mref{553} we get:
\be
\exp\ll(\int_{\infty_1}^{\infty_3}Z_1(y,\lambda)dy+\int_{\infty_2}^{z}Z_2(y,\lambda)dy-
\int_{\infty_3}^{z}Z_3(y,\lambda)dy\r)-1=0
\label{C3}
\ee

Using further the properties of $Z^\pm(z,\lambda)$ in the decomposition
$Z_k(z,\lambda)=Z^+(z,\lambda)+\sigma_kZ^-(z,\lambda)$ we get finally from \mref{C3}
\be
\exp\ll[\ll(\int^{\infty_3}_{z}+\int_{\infty_3}^{\infty_1}+\int_{\infty_1}^{\infty_2}+\int_{\infty_2}^{z}\r)
\ll(Z^+(y,\lambda)-Z^-(y,\lambda)\r)dy\r]-1=0
\label{C4}
\ee
where we have taken into account that $\int_{\infty_1}^{\infty_2}Z^+(y,\lambda)dy=
\int_{\infty_1}^{\infty_2}Z^-(y,\lambda)dy=0$.

However all the integrations in \mref{C4} run along paths none of which crosses any cut of the $C_{cut}$-plane, i.e. in the
connected domain of the plane. It then follows from the geometry of these paths that they form a closed contour in this
domain and therefore the integrations of $ Z^+(y,\lambda)$ and $Z^-(y,\lambda)$ along this contour have to vanish. But then
the condition \mref{C4} appears simply to be the identity.

A simple explanation of this a little bit unexpected situation is that in the considered domain $\psi_1(z,\lambda)$ grows
exponentially when $\Re z$ approaches the considered SL emerging from $z_2$. By comparison of $\psi_1(z,\lambda)=
W_{10}^{-\frac{1}{4}}(z)e^{\lambda\int_{z_1}^{z}\sqrt{W_{10}(y)}dy}\chi_{1}(z,\lambda)$ with its linear combination
\mref{C1} we have:
\be
\chi_{1\to 3}(\lambda)\chi_2(z,\lambda)-\chi_3(z,\lambda)=ie^{2\lambda\int_{z_5}^{z}\sqrt{W_{10}(y)}dy}
\label{C5}
\ee
where $\Re\ll(\lambda\int_{z_5}^{z}\sqrt{W_{10}(y)}dy\r)<0$ for $z\in T$ with $T$ being the strip on Fig.3 which boundaries
are formed by the SL's emerging from $z_1$ and $z_2$ from the one side and by the SL's emerging from $z_5$ from the other
side (there are no SL's inside $T$).

The relation \mref{C5} means that the l.h.s. of it has to vanish exponentially when $\lambda\to\infty$, i.e. the
semiclassical expansion of $\chi_{1\to 3}(\lambda)\chi_2(z,\lambda)-\chi_3(z,\lambda)$ has to vanish identically in $T$, i.e.
to any order.

On the other hand one never meets identities such as discussed above if $\psi_1(z,\lambda)$ is expressed by a linear
combination of a pair of FS's with different signatures.


\begin{thebibliography}{99}
\bibitem{11} Giller S., {\it J. Phys. A: Math. Theor.} {\bf 41} (2008) 465202
\bibitem{9} Hezari H., {\it International Mathematics Research Notices} {\bf 2008} (2008)
\bibitem{12} Eremenko A., Gabrielov A. and Shapiro B., {\it CMFT} {\bf 8} No. 2 (2008) 513-529
\bibitem{15} Martinez-Finkelstein A., Martinez-Gonzales P., Zarzo A., {\it J. Comp. App. Math.} {\bf 145} (2002) 167-182
\bibitem{14} Zelditch S., {\it Eur. Phys. J. Special Topics} {\bf 145} (2007) 271-286
\bibitem{13} Delabaere E., Dillinger H. and Pham F., {\it J. Math. Phys.} {\bf 38} (1997) 6126-6184
\bibitem{3} Giller S., {\it Acta Phys. Pol.} {\bf B21} (1990) 675-709
\bibitem{4}
\begin{enumerate}
\item Giller S., {\it J. Phys. A: Math. Gen.} {\bf 22} (1989) 647-661
\item Giller S., {\it J. Phys. A: Math. Gen.} {\bf 22} (1999) 2965-2990
\item Milczarski P. and Giller S., {\it J. Phys. A: Math. Gen.} {\bf 33} (2000) 357-393
\item Giller S., {\it Simple application of fundamental solution method in 1D quantum mechanics}, quant-ph/0107021
\item Giller S., {\it Acta Phys. Pol.} {\bf B35} (2004) 551-578
\end{enumerate}
\bibitem{5}
\begin{enumerate}
\item Giller S., {\it J. Phys. A: Math. Gen.} {\bf 21} (1988) 909-930
\item Giller S., {\it Acta Phys. Pol.} {\bf B23} (1992) 457-511
\item Giller S. and Milczarski P., {\it J. Phys. A: Math. Gen.} {\bf 32} (1999) 955-976
\item Giller S., {\it J. Phys. A: Math. Gen.} {\bf 33} (2000) 1543-1580
\item Giller S. and Milczarski P., {\it J. Math. Phys.} {\bf 42} (2001) 608-639
\end{enumerate}
\end{thebibliography}
\end{document}